\documentclass[12pt,intlimits,centertags,a4paper]{article}
\usepackage{amsmath}
\usepackage{nicefrac}
\usepackage{booktabs}
\usepackage{graphicx}
\usepackage{array}

\newcommand{\ar}{\textrm{arccot}}
\allowdisplaybreaks

\title{The Post-Newtonian Maclaurin Spheroids to
       Arbitrary Order}
 \date{Sept. 3 2003}
 \author{David Petroff \thanks{D.Petroff@tpi.uni-jena.de} }

\begin{document}

 \maketitle

 \begin{abstract}
  %%%%%%%%%%%%%%%%%%%%%
In this paper, we develop an iterative scheme to enable the explicit
calculation of an arbitrary post-Newtonian order for a relativistic body that
reduces to the Maclaurin spheroid in the appropriate limit. This scheme allows
for an analysis of the structure of the solution in the vicinity of bifurcation
points along the Maclaurin sequence. The post-Newtonian expansion is solved
explicitly to the fourth order and its accuracy and convergence are studied by
comparing it to highly accurate numerical results.
  %%%%%%%%%%%%%%%%%%%%%
 \end{abstract}
 
 \maketitle
 \section{Introduction} 
  %%%%%%%%%%%%%%%%%%%%%%%%%%%%
Upon the discovery of pulsars in 1968 and their identification as neutron
stars, it became apparent that a relativistic description of rapidly rotating,
compact stars was needed. Early work in this direction dealt with simplified
models for the matter making up these objects. In particular, Chandrasekhar
\cite{Chand67} looked at stars of constant density and calculated the first
post-Newtonian correction to the Maclaurin spheroids. Bardeen \cite{Bardeen71}
reexamined this work using a modified approach and gained new insight
regarding, foremost, the points of onset of secular, axisymmetric instability
along a one parameter Maclaurin curve.

Given the amount of work that has been done since then to study stars with more
realistic equations of state, the return to a model of constant density can
hardly be motivated by astrophysical considerations. Many other good arguments
however, suggest that precisely this model deserves closer attention: For one,
it allows, as we shall see, for the development of an iterative scheme to
calculate explicitly any order of the post-Newtonian expansion, limited in
practice only by computer algebra programs and the machines running them.
Furthermore, by considering an arbitrary order, one can study properties of the
full relativistic solution and carry out Bardeen's task of testing conjectures
``by going to higher orders in the relativistic expansion''. Finally, due to
the fortuitous circumstance of being in possession of highly accurate numerical
values, one can go even further. We are in the rare position of being able to
examine the behaviour of the post-Newtonian expansion itself, providing, by
analogy, a testbed for the most widely used analytic approximation within the
field of General Relativity.
  
In section~\ref{Remarks} of this paper, we motivate the method used here by
briefly describing Bardeen's approach \cite{Bardeen71} for the first order of
the expansion and explaining why modifications are necessary when going to
higher orders. Section~\ref{Basic} presents the line element and the Einstein
equations to be solved for iteratively. An iterative scheme allowing for the
explicit calculation of an arbitrary order is presented in section~\ref{Scheme}
and some properties of the solution are discussed in section~\ref{Properties}.
After providing by way of example the explicit calculation of a few expressions
and introducing various physical quantities, the PN approximation up to the
fourth order is compared with highly accurate numerical results in
section~\ref{Fourth}.
  %%%%%%%%%%%%%%%%%%%%%%%%%%%%

 \section{Preliminary Remarks}\label{Remarks}
  %%%%%%%%%%%%%%%%%%%%%%%%%%%
  For a given mass-density, $Q$, the Maclaurin spheroids and the relativistic
model both depend on two parameters.%
\footnote{In the Newtonian case, one of these is a mere scaling parameter.}
Since the post-Newtonian approximation describes the relativistic model in
terms of Newtonian parameters, some convention is needed to determine which
relativistic parameters are implied by the specification of the Newtonian ones.
Bardeen argued that the ``most appropriate choice'' compares Newtonian and
relativistic bodies of the same rest mass $M_0$ and angular momentum $J$ since
these quantities (together with $Q$) are coordinate independent and ``play the
primary role in the Hartle-Sharp \cite{HS} variation principle''. In this paper
we take a somewhat different approach since our purpose is less the comparison
of Newtonian and relativistic configurations, than the development of a method
for calculating an arbitrary order of the expansion. Therefore we use the
freedom that one has in defining the PN approximation in order to simplify the
mathematical structure of the equations. The remaining freedom regarding the
choice of a constant is left unspecified as long as possible. What effect the
specification of this constant then has, will be studied in section
\ref{Results} of the paper.

At this point, a brief description of the method that Bardeen used in
\cite{Bardeen71} will provide us with the basis for understanding the motivation
for the methods used in this paper.%  
\footnote{The reader who is interested in studying Bardeen's paper
 \cite{Bardeen71} may benefit from the following list of errata. The coefficient
 of $P_2(\eta)$ in eq.~(21) should read 
 $\frac{1}{p_2(\xi_s)}\left[\frac{1}{2}p_2(\xi)C_2 +
  \frac{{\xi_s}^2-\xi^2}{\xi_s(1+\xi_s)^{1/2}} \right]$.  The left hand side of
  eq.~(36) should read $\frac{1}{c^2}(v^2)_E(\xi_s,\eta)$ and on the right hand
  side $\frac{1}{12}D_2W_2(\eta)$. The $P_l(\eta)$ of eq.~(43) is to be replaced
  by $\left(P_l(\eta)-1 \right)$ and the second term in the first line of
  eq.~(55) by $+\frac{\Delta \xi}{c^2}\frac{\partial U}{\partial \xi}(\xi_s,1)$. }
Up to the first order of the PN approximation, one has to determine two metric
functions from Poisson-like equations as well as the unknown boundary of the
star. To solve for the metric functions, Bardeen used a Poisson-integral in
spheroidal coordinates $\xi$ and $\eta$, which represents potentials as an
expansion in terms of orthogonal polynomials in $\eta$. An iterative scheme for
the calculation of higher orders is only feasible if the sum over these
polynomials terminates. The conditions for the termination of the sum are that
the source remain a polynomial in $\eta$ and that the boundary of the star
remain a constant in $\xi$. Neither of these conditions is met with in
Bardeen's approach, which is why it is only appropriate up to the first order.
To that order it was possible to determine the metric functions in an elegant
way, because they can be decomposed into one piece containing the new
(post-Newtonian) source within the old boundary and another piece containing
the old source within the new boundary. For higher orders, such a procedure can
no longer be used and one has to devise a modified approach.

The approach used in this paper relies on the fact that an extended version of
the Poisson-integral is valid for Poisson-like equations even in modified
coordinates. Here coordinates will be introduced that are tailored to the
unknown boundary of the star and satisfy the condition that the boundary be a
constant in this coordinate. Furthermore we require that the unknown boundary of
the star when written as a function of the old coordinate $\xi$ be given as a
polynomial in $\eta$, a requirement that can be shown to be compatible with the
condition that the pressure vanish at the surface of the star. This requirement
ensures that the sources in the Poisson-like equations remain polynomials in
$\eta$. Thus we have to deal only with terminating sums to any order of the PN
approximation and the recursive method proposed here can be applied
indefinitely.
  %%%%%%%%%%%%%%%%%%%%%%%%%%

 \section{Basic Equations}\label{Basic}
  %%%%%%%%%%%%%%%%%%%
  The line element for an axially symmetric, stationary, asymptotically flat
space-time describing a perfect fluid with purely azimuthal motion can be
written in Lewis-Papapetrou coordinates as
 \[ ds^2=e^{2\mu}\left(d\varrho^2+d\zeta^2 \right) + 
       \varrho^2e^{2\lambda}\left(d\varphi - \omega\: dt \right)^2 -
       c^2 e^{2\nu}dt^2. \]
The metric functions $\mu$, $\lambda$, $\omega$ and $\nu$ depend only on
$\varrho$ and $\zeta$ and vanish at spatial infinity. The energy-momentum tensor
for the pressure $P$ and the mass density $Q$, which is merely the energy
density divided by $c^2$, is then given by
 \[ T_{\alpha\beta} = \left(Qc^2+P \right)u_\alpha u_\beta +g_{\alpha\beta}P, \]
where $Q$ is a constant up to the surface of the star. The matter of the star
rotates uniformly with an angular velocity $\Omega$. We introduce the
spheroidal coordinates
 \[ \varrho^2={a_0}^2(1+\xi^2)(1-\eta^2)\ \ \ \textrm{and}\ \ \ \zeta=a_0\xi\eta,
  \ \ \ \eta\in[-1,1],\ \ \xi \in [0,\infty), \]
and obtain from various combinations of the Einstein equations the following
partial differential equations for the metric functions (with $G=1$ for the
gravitational constant):
\begin{subequations}\label{Feldgl}
 \begin{align}
  \begin{split}
  \Delta_2 \nu =&\, \frac{4 \pi e^{2\mu}}{c^4}
                  \left[\frac{1+\tilde{v}^2}{1-\tilde{v}^2}(Qc^2+P)+2P \right]
                  -{\cal L}(\nu,\nu+\lambda)  \\
                &\, +\frac{1}{2}\tilde{\Omega}^2(1+\xi^2)(1-\eta^2) 
                    e^{2\lambda-2\nu}{\cal L}(\tilde{\omega},\tilde{\omega}),
                    \label{nuGl}
  \end{split} \\
  \Delta_3(\lambda+\nu) =&\, \frac{16\pi e^{2\mu}}{c^4}P 
                   - {\cal L}(\nu+\lambda,\nu+\lambda), \label{lambdaGl} \\  
  \Delta_4 \tilde{\omega} =&\, \frac{-16\pi 
                 (1-{\tilde \omega})e^{2\mu}}{c^4(1-\tilde{v}^2)}
                 (Qc^2+P) - {\cal L}({\tilde  \omega},3\lambda-\nu),
                 \label{omegaGl} \\
  \begin{split}		     
  \Delta_1 \mu =&\, \frac{-4\pi e^{2\mu}}{c^4}(Qc^2+P)+{\cal L}(\nu,\lambda)+
                  \frac{1}{4}(1+\xi^2)(1-\eta^2)e^{2\lambda-2\nu}
                     {\cal L}(\tilde{\omega},\tilde{\omega})  \\ 
              &\, + \frac{1}{a_0^{\,2}(\xi^2+\eta^2)}
                    \left(\xi\nu_{,\xi}-\eta\nu_{,\eta} \right).
                    \label{muGl}
  \end{split}              
 \end{align}
\end{subequations}
The differential operators in the above equations are defined by
\begin{align*}
 {\cal L}(\phi,\chi) :=&\, \left[(1+\xi^2)\phi_{,\xi}\chi_{,\xi} 
                                 +(1-\eta^2)\phi_{,\eta}\chi_{,\eta}
                           \right]/a_0^{\,2}(\xi^2+\eta^2)
 \\ \intertext{and}
  \Delta_m:=&\, \left[
     (1+\xi^2)\frac{\partial^2}{\partial\xi^2} 
    +(1-\eta^2)\frac{\partial^2}{\partial\eta^2}
    +m\xi\frac{\partial}{\partial\xi} - m\eta\frac{\partial}{\partial\eta}
  \right] /a_0^{\,2} (\xi^2 + \eta^2)
 \end{align*}
and the dimensionless function in eq.~(\ref{omegaGl}) by
\[ \tilde{\omega}:=\frac{\omega}{\Omega}. \]
Note that the operator $\Delta_2$ is simply the Laplace operator in a flat
three-dimensional space. The dimensionless pressure $\tilde{P}:=P/Qc^2$ is
related to the metric functions by
\begin{align}\label{DruckGl}
 \sqrt{1+\tilde{v}^2}\,(1+\tilde{P})\, e^\nu = \textrm{const.} = 1- \gamma\\
 \intertext{with}
   \tilde{v} := \varrho\tilde{\Omega} (1-\tilde{\omega}) e^{\lambda-\nu}/a_0
   \ \ \ \textrm{and}\ \ \ \tilde{\Omega}:= a_0 \Omega/c. \nonumber
\end{align}
  %%%%%%%%%%%%%%%%%%

 \section{The Iterative Scheme}\label{Scheme}
  \subsection{The Expansion}\label{Expansion}
   %%%%%%%%%%%%%%%%%
   The system of partial differential equations~(\ref{Feldgl}) is simplified
by expanding the relevant quantities in terms of a dimensionless relativistic
parameter. Here we choose the square root of the parameter%
\footnote{The square root was chosen in order to enable a more convenient indexing of
the expansion coefficients.}
used in \cite{Bardeen71} defined by
\begin{equation}\label{ParameterGl}
 \varepsilon^2 := \frac{8 \pi Q {a_0}^2 \xi_\text s \sqrt{1+{\xi_\text s}^2}}{3c^2}.
\end{equation}
The three variables entering into this definition completely specify the
Newtonian Maclaurin spheroid. $Q$ is the mass density, $a_0$ the focus of the
ellipse describing the surface of the star in cross-section and $\xi_\text s$ the
value of the surface's $\xi$ coordinate. These are the same quantities which
will enter into the PN expansion, but the latter two lose their simple
geometrical meaning. The parameter $\varepsilon$ remains finite in both the
spherical limit, given by $\xi_\text s \to \infty$ and $a_0 \propto 1/\xi_\text s$ and the
disc limit, which is given by $\xi_\text s \to 0$ and $Q \propto 1/\xi_\text s$ for
non-vanishing mass.

The expansion of the dimensionless metric functions and the constants reads as
follows:
\begin{equation}\label{Entwicklung}
 \begin{aligned}
  \nu&=\sum\limits_{n=2}^\infty \nu_n \varepsilon^n &
  \lambda&=\sum\limits_{n=2}^\infty \lambda_n \varepsilon^n &
  \tilde{\omega}&=\sum\limits_{n=2}^\infty \tilde{\omega}_n \varepsilon^n \\
  \mu&=\sum\limits_{n=2}^\infty \mu_n \varepsilon^n &
  \gamma&=\sum\limits_{n=2}^\infty \gamma_n \varepsilon^n & 
  \tilde{\Omega}&=\sum\limits_{n=1}^\infty \tilde{\Omega}_n \varepsilon^n.
 \end{aligned}
\end{equation}
As was already mentioned, $Q$ is held constant to any order of the
approximation, which is why it does not appear in eq.~(\ref{Entwicklung}) and
any other quantities of interest, such as $\tilde{v}$ or $\tilde{P}$ can be
expressed in terms of these six quantities.

If these expansions are substituted into eqs~(\ref{Feldgl}), then comparing
coefficients of $\varepsilon$ yields differential equations for the metric
functions of the form $\Delta_m\phi_i=F$, where $\phi=\nu, \lambda,
\tilde{\omega}, \mu$. Because the right hand side of
eqs~(\ref{nuGl}--\ref{omegaGl}) depends only on $\phi_{i-j}$, $j>0$, one can
solve for $\phi_i$ if the lower order functions are already known. In the case
of $\mu_i$, one can calculate it from eq.~(\ref{muGl}) after having determined
the other three functions to this order, or one can compute it from an integral
over $\eta$.

Because an analytic solution, the Maclaurin solution, for the first step is
known, these equations would provide an iterative process for the determination
of the metric functions to any order if the shape of the star were known. The
boundary of the star also has to be determined iteratively however.

We represent the surface of the star by the equation,
\begin{equation}\label{OberGl}
 \xi=\xi_\text B(\eta)=\xi_\text s \left(1+\sum_{j=0}^{k} \sum_{k=2}^\infty
  S_{jk}\,C_j^{1/2}(\eta)\,\varepsilon^{k} \right)
  \equiv \xi_\text s \left(1+\sum_{k=2}^\infty
                       B_k(\eta) \varepsilon^{k} \right), 
\end{equation}
where we have already taken into account that the boundary is an ellipsoid
$\xi=\xi_\text s$ in the Newtonian order. We also require that the sum over the
Legendre polynomials $C_j^{1/2}(\eta)$, a special case of the Gegenbauer
polynomials discussed in section~\ref{Poisson}, terminate and show in
section~\ref{Gestalt} that this leads to a consistent solution.
   %%%%%%%%%%%%%%%%%%
  \subsection{Solving the Poisson-like Equations}\label{Poisson}
   %%%%%%%%%%%%%%%%%%
In the last section an iterative scheme was proposed for the determination of
the metric functions in which an equation of the form $\Delta_m\phi=F$ need be
solved for a known function $F=F(\xi,\eta)$. The regular and asymptotically flat
solution of this equation is given by
\begin{equation}\label{IntformelGl}
 \begin{aligned} 
  \phi(\xi,\eta) &=\, 
             a_0^{\,2} \sum\limits_{l=0}^\infty K_l^m
              C_l^{\frac{m-1}{2}}\!(\eta)\,\times \\
              &\;\; \Big[h_l^m(\xi)\int\limits_0^\xi\int\limits_{-1}^1         
                g_l^m\!(\xi')C_l^{\frac{m-1}{2}}\!(\eta')F\!(\xi',\eta')        
                k_m\!(\xi',\eta')\, d\eta'\, d\xi' + \\
              &\;\; g_l^m(\xi)\int\limits_\xi^\infty\int\limits_{-1}^1         
                h_l^m\!(\xi')C_l^{\frac{m-1}{2}}\!(\eta')F\!(\xi',\eta')        
                k_m\!(\xi',\eta')\, d\eta'\, d\xi' \Big].                       
 \end{aligned}
\end{equation}
In the above equation $C_i^j$ are the Gegenbauer polynomials, $g_i^j$ and
$h_i^j$ are two linearly independent solutions of the (homogeneous) Gegenbauer
equation defined by
\begin{align}\label{ghGl}
 \begin{split}
   g_l^m(\xi):=&\, C_l^{\frac{m-1}{2}}(i\xi)\\
   h_l^m(\xi):=&\, g_l^m(\xi) \int\limits_\xi^\infty \frac{d\xi'}{\left( g_l^m(\xi')
   \right)^2  E(\xi')}\ \ \ \ (l,m) \neq (0,1)\ \ \textrm{and}\\
   h_0^1(\xi):=&\, \textrm{arcsinh}(\xi)
 \end{split} \\
 \intertext{with}
  E(\xi):=&\, \exp\left( \int_0^\xi
    \frac{m\xi'}{1+\xi'^2}\,  d\xi' \right) = \left( 1+\xi^2 \right)^{m/2}.
  \nonumber
\end{align}
The term
\begin{equation}\label{km}
  k_m(\xi,\eta)\,d\eta\,d\xi:= \left[\left(1+\xi^2\right)\left(1-\eta^2\right)\right]^
                 {\frac{m}{2}-1}(\xi^2+\eta^2)\,d\eta\,d\xi
\end{equation}		     
is a product of the volume element and the appropriate weight function for the
Gegenbauer polynomials and
\begin{eqnarray*}
 K_l^m &=& \frac{l! \left(l+\frac{m}{2} -\frac{1}{2} \right)
             \left[ \Gamma\!\left(\frac{m}{2} -\frac{1}{2} \right) \right]^2}
             {\pi 2^{2-m} \Gamma\!\left(m-1+l \right)}\ \ {\rm for}\ m>1 \\
 K_l^1 &=& \frac{l^2}{2\pi}\ \ {\rm for}\ l>0\ \ {\rm and}\\ \\
 K_0^1 &=& \frac{1}{\pi} 
\end{eqnarray*}
are normalizing constants.

In eq.~(\ref{IntformelGl}) the integrands jump at the surface of the star
because of the jump in the mass density. It is therefore necessary to split
them into integrals over the interior and exterior of the star. Clearly if the
surface of the star is given, as with the leading order, by a constant
$\xi=\xi_\text s$, then this division is trivial. If the boundary depends on $\eta$,
then matters are complicated considerably. As of the second order in the
expansion, the $\eta$-integrals no longer run over the interval $\eta \in
[-1,1]$ meaning that one can no longer make use of the orthogonality of the
Gegenbauer polynomials and one is faced with non-terminating sums. We hence
introduce new coordinates in order to circumvent these difficulties.
   %%%%%%%%%%%%%%%%%%%
  \subsection{New Coordinates}
   %%%%%%%%%%%%%%%%%%%%
   We introduce the coordinates
\begin{equation}\label{KoordGl}
 \psi=\frac{\xi_\text s\xi}{\xi_B(\eta)},\ \ \eta=\eta
\end{equation}
implying that $\psi=\xi_\text s$ is the boundary of the star, cf. 
eq.~\eqref{OberGl}. The new coordinate $\psi$ is a function of both $\eta$ and
$\varepsilon$ and contains the unknown coefficients $S_{jk}$. We rewrite
eqs~(\ref{Feldgl}) in terms of the new coordinates and manipulate them such
that the left hand side has the same form as beforehand, but with $\psi$
replacing $\xi$. For example, the equation for $\nu=\nu(\psi,\eta)$ now reads
\[ \left[
     (1+\psi^2)\frac{\partial^2\nu}{\partial\psi^2} 
    +(1-\eta^2)\frac{\partial^2\nu}{\partial\eta^2}
    +2\psi\frac{\partial\nu}{\partial\psi} - 2\eta\frac{\partial\nu}{\partial\eta}
  \right] /a_0^{\,2} (\psi^2 + \eta^2)=\bar{F}.\]
These new field equations are again expanded%
\footnote{Note that the coefficients $\phi_i(\psi,\eta)$ are not mere
transformations of $\phi_i(\xi,\eta)$ since $\psi$ depends on $\varepsilon$.
Thus one must substitute $\psi$ of eq.~\eqref{KoordGl} into $\sum_{j=2}^n
\phi_j(\psi,\eta)$ and expand the result in terms of $\varepsilon$ in order to
express $\phi_k$, $k\leq n$, in terms of $\xi$.}
in terms of $\varepsilon$ in order to obtain a system of equations for $\phi_i$
as was explained in section~\ref{Expansion}. Since $\psi=\xi+{\cal
O}(\varepsilon^2)$ holds, the new equations for $\phi_i(\psi,\eta)$ also depend
only on known functions, thereby enabling their recursive determination. 

The derivation of eq.~(\ref{IntformelGl}) relies on the fact that in the
coordinates $(\xi,\eta)$, the line $(0,\eta)$ is identical to the line
$(0,-\eta)$ and that at spatial infinity we have $\xi \to \infty$. These
properties hold for $\psi$ as well and an analysis of the derivation shows
that we are free to use eq.~\eqref{IntformelGl} as it stands, only replacing $\xi$
by $\psi$.

In changing coordinates we have mapped the star onto the rectangle $[0,\xi_\text s]
\times [-1,1]$, which means that the division of the integrals into inner and
outer domains is trivial. The price that one pays for the simplicity in the
structure of the integrals is that the sources of the Poisson-like equations
become quite unwieldy. But the exchange of a conceptual for a mechanical
difficulty can be termed a good deal, and all the more so when its result is
the facilitation of the whole scheme.
   %%%%%%%%%%%%%%%%%
  \subsection{Determining the Shape of the Star}\label{Gestalt}
   %%%%%%%%%%%%%%%
   Due to the factor $c^2$ in $g_{tt}$ of the line element, it is necessary to
determine the function $\nu_{i+2}$ in order to calculate the
$i^\text{th}$ order of the PN approximation. This is the only metric
function that depends on the unknown coefficients $S_{ji}$ of the star's
boundary.%
\footnote{The other metric functions depend only on $S_{jk}$, with $j<i$.}
To determine these coefficients, one calculates the pressure from
eq.~(\ref{DruckGl}) along the boundary of the star and sets the coefficients of
an expansion in terms of $\eta$ equal to zero. In discussing the boundary,
however, it turns out to be useful to leave a portion of the Poisson-integral
for $\nu_{i+2}$ unevaluated in order to arrive at the integral equation
\begin{equation}\label{Intformel}
 {\tilde{\Omega}_1}^2 (1-\eta^2)\,{\xi_\text s}^2\,B_i(\eta) =
   \sum_{l=0}^\infty K_l^2\,C_l^{1/2}(\eta)
   \int_{-1}^1 C_l^{1/2}(\tilde{\eta}) f_l(\tilde{\eta})
     B_i(\tilde{\eta})\,d\tilde{\eta} +  b_i(\eta).
\end{equation}
The function $B_i$ is defined in  eq.~(\ref{OberGl}) and contains the entire
dependence on $S_{ji}$. The function $b_i(\eta)$ is short for the remaining
terms that result from eq.~\eqref{DruckGl} and is a known polynomial of order
$i+2$. The function $f_l(\tilde{\eta})$ is given by 
\begin{align*}
 f_l(\tilde{\eta}):=
  &g_l^2(\xi_\text s) \int_{\xi_\text s}^\infty h_l^2(\psi) 
       \Big[ 2\tilde{\eta} \psi (\dot{\nu}_2^\text o)'
           -\psi(1-\tilde{\eta}^2)(\dot{\nu}_2^\text o)'' + 2\ddot{\nu}_2^\text o  
            -l(l+1)\psi\dot{\nu}_2^\text o\Big]\,d\psi
 \\		
 + &h_l^2(\xi_\text s) \int_0^{\xi_\text s} g_l^2(\psi) 
       \Big[ 2\tilde{\eta} \psi (\dot{\nu}_2^\text i)'
           -\psi(1-\tilde{\eta}^2)(\dot{\nu}_2^\text i)'' + 2\ddot{\nu}_2^\text i 
           + \frac{3\psi^2}{\xi_\text s\sqrt{1+{\xi_\text s}^2}}-l(l+1)\psi\dot{\nu}_2^\text
	     i 
       \Big]\,d\psi, 		
\end{align*}
where a dot and prime indicate  partial derivatives with respect to $\psi$ and
$\tilde{\eta}$ respectively and the superscripts `i' and `o' refer to the
regions inside and outside the star. Since $f_l(\eta)$ is a polynomial of second order,
the sum in eq.~(\ref{Intformel}) terminates for polynomial $B_i(\eta)$. Indeed,
for the form of $B_i$ chosen in eq.~(\ref{OberGl}), one arrives at a system of
$i+2$ algebraic equations for $i+3$ unknowns (there are $i+1$ $S_{ij}$ to
determine as well as $\tilde{\Omega}_{i+1}$ and $\gamma_{i+2}$).%
\footnote{We shall see in section~\ref{Reflection} that $S_{ij}=0$ for odd $j$.}
We choose to use this system to determine all these constants but for
$\gamma_{i+2}$. As mentioned in section \ref{Remarks}, this last constant can
be chosen arbitrarily, which amounts to specifying ``which'' PN-approximation
one wishes to have, i.e. which relativistic body is to be associated with a
given Maclaurin spheroid. The choice of $\gamma_{i+2}$ will be discussed further
in section \ref{Results}.

We have shown that the form chosen for the surface of the star is consistent
with the Einstein equations to any order of the PN expansion. This is not to say
that this choice is unique. One can easily see that the form chosen in
\cite{Bardeen71} is incompatible with that chosen here, since it is not a
polynomial in $\eta$. There the surface was derived having stipulated that the
`generating' Maclaurin spheroid should have the same rest mass and angular
momentum as the PN star, a condition that cannot be satisfied with the approach
chosen here. In lieu of the freedom to choose two constants, we have chosen a
form for the boundary of the star most appropriate to our goal of devising an
iterative scheme and can choose only one further constant.
   %%%%%%%%%%%%%%%%
  
  \section{Properties of the Solution}\label{Properties}
   %%%%%%%%%%%%%%%%
   \subsection{Reflectional Symmetry}\label{Reflection}

In Newtonian physics, it is known, that stationary, axisymmetric bodies are
necessarily symmetric with respect to a reflection through the $\zeta=0$ plane
(see e.g. \cite{Lichtenstein}). Although authors (e.g. \cite{Lindblom}) have
speculated that the same holds in General Relativity, it has not yet been
proved true. In the case considered here, this symmetry arises automatically. A
function $f$ exhibits reflectional symmetry in $\xi$-$\eta$ (or $\psi$-$\eta$)
coordinates precisely when it is an even function of $\eta$. Because of the
orthogonality of the Gegenbauer polynomials, the terms in the sum of
eq.~(\ref{IntformelGl}) for odd $l$ are zero if $F$ is a polynomial in
$\eta^2$, a condition which turns out to be satisfied. The odd terms, which are
provided by the unknown boundary coefficients $S_{li}$ must be zero for the
boundary condition to be fulfilled. Thus we have shown that any axially
symmetric, stationary, relativistic solution that is continuously connected to
the Maclaurin spheroids is symmetric with respect to reflections through the
$\zeta=0$ plane.

\subsection{Powers of the Relativistic Parameter}

Consideration of the field equations together with the knowledge of the
Newtonian behaviour of the dimensionless metric functions, shows that their
expansion coefficients $\phi_i$ begin with $i=2$ and are non-zero only for even
$i$. The same holds naturally for $\gamma_i$, whereas $\tilde{\Omega}_j$ begins
with $j=1$ and appears only with odd powers powers of $j$. Because of the
choice to work with the dimensionless functions introduced here, it is most
appropriate to refer to the $n^\text{th}$ order of the PN approximation
and not the half orders in between. What we mean by the $n^\text{th}$
order is that the quantities $\lambda$, $\tilde{\omega}$, $\mu$ and $\xi_B$ are
expanded up to and including the order ${\cal O}(\varepsilon^{2n})$,
$\tilde{\Omega}$ up to ${\cal O}(\varepsilon^{2n+1})$ and $\nu$ and $\gamma$
to  ${\cal O}(\varepsilon^{2n+2})$.

%In the remainder of this paper, the properties mentioned here and in the
%preceding section will be taken into account. Thus statements regarding
%$S_{ij}$, for example, hold for even values of $i$ and $j$.

\subsection{Singularities in Parameter Space}      

By comparing the highest coefficient of $\eta$ in eq.~(\ref{Intformel}) one can
arrive at the equation
\begin{equation}\label{Sii}
 S_{ii} = \frac{t_i}{{\xi_\text s}^2 \tilde{\Omega}_1^2 - a_{i+2}}
\end{equation}
with
\[
t_i \propto \int_{-1}^1 C_{i+2}^{\nicefrac{1}{2}}(\bar{\eta}) b_i(\bar{\eta})\,
d\bar{\eta}\]
and
\[
C_{i+2}^{\nicefrac{1}{2}}(\eta) f_{i+2}(\eta)=:\sum_{n=0}^1 \bar{a}_{2n,i+2}\eta^{2n},\ \ 
a_{i+2}:=\bar{a}_{2,i+2}\]
and where $f_{i+2}$ and $b_i$ are defined in eq.\eqref{Intformel}.
It can be shown that the denominator of eq.~(\ref{Sii}) is proportional to the
expression
\begin{equation}\label{Sing}
 g_{i+2}^{2}(\xi_\text s) h_{i+2}^{2}(\xi_\text s) - \xi_\text s(1-\xi_\text s
 \textrm{arccot}(\xi_\text s))=:G_i(\xi_\text s).
\end{equation} 
For a given (even) $i$, this expression vanishes for precisely one value of
$\xi_\text s$, let us say for $\xi_\text s=\xi_{i+2}^*$. These values,
beginning with $i=2$, are the points of onset of axisymmetric, secular
instability and the bifurcation points of new axisymmetric solutions, see
\cite{Bardeen71,HuE,AKM2}, and numerical values for the first few of them can be
found in Table~\ref{epsstar}.
\begin{table}
 \centerline{
  \begin{tabular}{cccc} \toprule
  $l$ & $\xi_{2l}^*$ & $e_{2l}^*$ &    $r_\mathrm{p}/r_\mathrm{e}$ \\ \midrule
   2   &  0.17383011  & 0.98522554  & 0.17126187\\
   3   &  0.11230482  & 0.99375285  & 0.11160323\\
   4   &  0.08303471  & 0.99657034  & 0.08274493\\
   5   &  0.06588682  & 0.99783651  & 0.06574427\\  
  \bottomrule 
 \end{tabular} }
 \caption{Numerical values for $\xi_{2l}^*$ and the corresponding Newtonian
 eccentricities and ratios of polar to equatorial radii given by
 $e=1/\sqrt{1+{\xi_\text s}^2}$ and $r_{\text p}/r_{\text e}=\xi_{\text
 s}/\sqrt{1+{\xi_{\text s}}^2}$.  \label{epsstar}}
\end{table}
Since $t_i$ of eq.~\eqref{Sii} is not zero at the point $\xi_\text
s=\xi_{i+2}^*$, these bifurcation points are singularities in the two
dimensional parameter space ($\xi_\text s$,$a_0$) or ($\xi_\text
s$,$\varepsilon$). For values of $\xi_\text s$ differing only slightly from
$\xi_{i+2}^*$, the PN configurations have properties similar to those of the
Newtonian configurations that branch off from the Maclaurin sequence at these
points. The Maclaurin configuration itself cannot be reached for bodies with
non-zero mass however, and even neighbouring configurations have strict mass
limitations, since $\varepsilon$ must be made very small in order that the PN
series converge. This mass limitation can be inferred, for example, by referring
to the tables in Appendix~\ref{tables}. Because the $n^{\text th}$ PN order
possesses a pole of order $2n-1$ at the point $\xi=\xi_4^*$, we expect the
coefficients for the expansion to grow large in the vicinity of this point.%
\footnote{A lengthier discussion regarding the order of the poles at
$\xi_{2i+2}^*$, $i>1$ can be found in \cite{P}.}
This is indeed the case as can be seen in these tables by referring to the row
with $\xi_{\text s}=0.17$. The series containing these coefficients converge
only for sufficiently small $\varepsilon$ as indicated above.   
   %%%%%%%%%%%%%
   
  \section{Explicit Solution to the Fourth Order}\label{Fourth}
   \subsection{The Metric Functions and the Constants}
    %%%%%%%%%%
Using the iterative scheme described above, the four metric functions and the
constant $\tilde{\Omega}$ were explicitly solved up to the fourth
post-Newtonian order. These calculations could in principle be carried out {\it
ad infinitum}, but the lengthiness of the expressions (the fourth order
functions would fill several hundred pages) puts a practical limit on the order
that can be determined. Here we will merely carry out, by way of example, the
calculation of the first few terms.

The expansion of eq.~(\ref{nuGl}) with respect to the relativistic parameter
$\varepsilon$ yields the Newtonian equations
\begin{subequations}\label{nu2PoissonGl}
 \begin{align}
  \Delta_2\nu_2^\text i &= \frac{4\pi}{\varepsilon^2 c^2}Q = 
                 \frac{3}{2 a_0^{\,2}\xi_\text s\sqrt{1+\xi_\text s^{\,2}}}\\
  \intertext{for the interior region ($\xi<\xi_\text s$) and}
  \Delta_2\nu_2^\text o &= 0.
 \end{align}
\end{subequations}
for the region $\xi>\xi_\text s$ exterior to the body. These equations are solved
using eq.~(\ref{IntformelGl}) to obtain
\begin{subequations}\label{nu2Gl}
 \begin{align}
  \begin{split} \nu_2^\text i =& 
  %%%%%%%%%%%%%
  -\frac{1}{2\,\xi_\text s\,\sqrt {1+{\xi_\text s}^{2}}}\\ 
&\bigg\{ \Big[ 
 \xi_\text s(1+{\xi_\text s}^2)h_0^2(\xi_\text s)+\frac{1}{2}( {\xi_\text s}^2-\xi^2)\Big] 
C_0^{1/2}(\eta)\\ 
&+\Big[
 -\xi_\text s\,(1+{\xi_\text s}^{2})\, g_2^2(\psi) \,h_0^2(\xi_\text s)\\
& +\frac{1}{3}\left( (3\,{ \xi_\text s}^{2}+2)g_2^2(\psi)+1\right)
\Big]C_2^{1/2}(\eta) \bigg\}
  %%%%%%%%%%%%
  \end{split}
  \\ \intertext{and}  
  \begin{split} \nu_2^\text o =& 
  %%%%%%%%%
  \frac{-\sqrt {1+{\xi_\text s}^{2}}}{2} \left( h_0^2(\psi)\,C_0^{1/2}(\eta)
-h_2^2(\psi)\,C_2^{1/2}(\eta)
 \right) 
  %%%%%%%%%%%%%
  \end{split}
 \end{align}
\end{subequations}
(see Appendix~\ref{gh} for a list of the first few $g_l^m$ and $h_l^m$). One
can verify that $\nu_2^\text i(\xi_\text s,\eta)=\nu_2^\text o(\xi_\text s,\eta)$ holds. The
requirement that the pressure vanish to this order of the expansion then fixes
the two remaining constants:
\begin{equation}\label{KonstGl}
 \tilde{\Omega}_1= -\frac{3}{2\sqrt{1+{\xi_\text s}^2}}h_2^2(\xi_\text s)
   \ \ \ \ \mathrm{and}\ \ \ \ 
   \gamma_2 = \frac{\sqrt{1+{\xi_\text s}^2}}{2}\left(h_0^2(\xi_\text s)-h_2^2(\xi_\text s)
   \right).
\end{equation}
Expanding eqs~(\ref{lambdaGl}) and (\ref{muGl}) shows that 
$\lambda_2=\mu_2=-\nu_2$ holds and one need only expand eq.~(\ref{omegaGl})
to obtain the last of the functions $\phi_2$, where
$\phi=\nu,\lambda,\tilde{\omega},\mu$. This expansion leads to the equations
\begin{align*}
 \Delta_4 \tilde{\omega}_2^\text i &= \frac{-6}{a_0^{\,2}\,\xi_\mathrm{s} \sqrt{1+\xi_\mathrm{s}^{\,2}}}
 \intertext{and}
 \Delta_4 \tilde{\omega}_2^\text o &= 0,
\end{align*}
with the solutions
\begin{subequations}\label{omegas2Gl}
 \begin{align}
   \begin{aligned} \tilde{\omega}_2^\text i =& 
   %%%%%%%%%%%%%%
   \frac{1}{5\,\xi_\mathrm{s}\sqrt {1+{\xi_\text s}^{2}}}\\
& \bigg\{\Big[
 6\xi_\mathrm{s}\left( 1+{\xi_\text s}^{2} \right) ^{2}h_0^4(\xi_\mathrm{s})
+3\,\left( {\xi_\mathrm{s}}^2-\xi^2 \right)\Big] C_0^{\nicefrac{3}{2}}(\eta)\\
&+ \Big[ -\xi_\mathrm{s}\,\left( 1+{\xi_\text s}^{2} \right)^{2}
g_2^4(\xi)\,h_0^4(\xi_\mathrm{s})\\
&+{\frac {1}{15}}\,\left((5\,{\xi_\text s}^{2}+4)\,g_2^4(\xi)
   +6\right)\Big] C_2^{\nicefrac{3}{2}}(\eta) \bigg\}
   %%%%%%%%%%%%%%
   \end{aligned}
   \\ \intertext{and}
   \begin{aligned} \tilde{\omega}_2^\text o =& 
   %%%%%%%%%%%%%%%%
   \frac{6}{5}\left( 1+{\xi_\text s}^{2} \right) ^{\nicefrac{3}{2}}
\left(h_0^4(\xi)\,C_0^{\nicefrac{3}{2}}(\eta)-
      h_2^4(\xi)\,C_2^{\nicefrac{3}{2}}(\eta)\right)
   %%%%%%%%%%%%%%%%%%
   . \end{aligned}
 \end{align}
\end{subequations}

Using the scheme proposed here, the calculation of the higher orders is much
lengthier than the calculations just shown, but otherwise identical. For $\nu_4$
for example, the last metric function needed in describing the first
post-Newtonian correction, we find
\begin{subequations}\label{nu4QuellGl}
 \begin{align}
  \begin{aligned} {a_0}^2\Delta_2 \nu_4^\text i =&\, 
  %%%%%%%%%%%%%
  \Bigg[{\displaystyle \frac {3}{4}} \,{\displaystyle \frac {( - 1 - 9\,
\eta ^{4} + 63\,\eta ^{4}\,\psi ^{2} + 6\,\eta ^{2} - 54\,\psi ^{
2}\,\eta ^{2} + 3\,\psi ^{2})\,\sqrt{1 + {\xi_\text s}^{2}}\,
\mathrm{arccot}({\xi_\text s})}{\eta ^{2} + \psi ^{2}}} \\ &\,
 - {\displaystyle \frac {3}{4}}  
(63\,\psi ^{2}\,{\xi_\text s}^{2}\,\eta ^{4} + 42\,\eta ^{4}\,\psi 
^{2} - 6\,\eta ^{4} - 9\,{\xi_\text s}^{2}\,\eta ^{4} - 54\,\eta ^{2
}\,\psi ^{2}\,{\xi_\text s}^{2} + 6\,\eta ^{2}\,{\xi_\text s}^{2} \\ &\,
\mbox{} + 2\,\eta ^{2} - 36\,\psi ^{2}\,\eta ^{2} - {\xi_\text s}^{2
} + 3\,\psi ^{2}\,{\xi_\text s}^{2} + 2\,\psi ^{2}) \left/ {\vrule 
height0.54em width0em depth0.54em} \right. \!  \! \Big((\eta ^{2} + 
\psi ^{2})\,{\xi_\text s} 
\sqrt{1 + {\xi_\text s}^{2}}\Big)\Bigg]{S_{22}}
\\ &\, 
+ \left[  \!  - {\displaystyle \frac {3}{2}} \,{\displaystyle 
\frac {(3\,\eta ^{2} - 1)\,\sqrt{1 + {\xi_\text s}^{2}}\,\mathrm{
arccot}({\xi_\text s})}{\eta ^{2} + \psi ^{2}}}  + {\displaystyle 
\frac {3}{2}} \,{\displaystyle \frac {2\,\eta ^{2} + 3\,\eta ^{2}
\,{\xi_\text s}^{2} + 2\,\psi ^{2} - {\xi_\text s}^{2}}{(\eta ^{2} + 
\psi ^{2})\,{\xi_\text s}\,\sqrt{1 + {\xi_\text s}^{2}}}}  \!  \right] 
 {S_{02}} 
\\ &\, - {\displaystyle \frac {21}{4}} \,
{\displaystyle \frac {(1 + \psi ^{2})\,( - 1 + \eta )\,(\eta  + 1
)\,{\tilde{\Omega} _{1}}^{\;2}}{{\xi_\text s}\,\sqrt{1 + {\xi_\text s}^{2}}}}  - 
{\displaystyle \frac {9}{2}} \,{\displaystyle \frac {{\gamma _{2}
}}{{\xi_\text s}\,\sqrt{1 + {\xi_\text s}^{2}}}}  \\ &\,
\mbox{} + {\displaystyle \frac {45}{16}} \,{\displaystyle \frac {
(1 + 3\,\psi ^{2}\,\eta ^{2} - \psi ^{2} + \eta ^{2})\,\mathrm{
arccot}({\xi_\text s})}{{\xi_\text s}}}  \\ &\,
\mbox{} - {\displaystyle \frac {45}{16}} \,{\displaystyle \frac {
 - \psi ^{2}\,{\xi_\text s}^{2} + \eta ^{2}\,{\xi_\text s}^{2} - {\xi 
_{s}}^{2} + 3\,\eta ^{2}\,\psi ^{2}\,{\xi_\text s}^{2} + 2\,\psi ^{2
}\,\eta ^{2}}{{\xi_\text s}^{2}\,(1 + {\xi_\text s}^{2})}}
  %%%%%%%%%%%%%%
  \end{aligned} 
 \\ \intertext{and}
  \begin{aligned}  {a_0}^2\Delta_2 \nu_4^\text o =&\, 
  %%%%%%%%%%%%%%%%%
  %% Created by Maple 8.00 (IBM INTEL LINUX)
%% Source Worksheet: Untitled (4)
%% Generated: Fri Jul  4 09:43:58 2003
\Bigg[{\displaystyle \frac {3}{4}} \,{\displaystyle \frac {( - 1 - 9\,
\eta ^{4} + 63\,\eta ^{4}\,\psi ^{2} + 6\,\eta ^{2} - 54\,\psi ^{
2}\,\eta ^{2} + 3\,\psi ^{2})\,\sqrt{1 + {\xi_\text s}^{2}}\,
\mathrm{arccot}(\psi )}{\eta ^{2} + \psi ^{2}}} 
\\ &\, - {\displaystyle \frac {3}{4}}  
\psi (27\,\eta ^{4} + 96\,\eta ^{4}\,\psi ^{2} + 63\,\eta ^{4}\,
\psi ^{4} - 16\,\eta ^{2} - 78\,\psi ^{2}\,\eta ^{2} - 54\,\eta 
^{2}\,\psi ^{4} \\ &\,
\mbox{} + 2\,\psi ^{2} - 3 + 3\,\psi ^{4})\sqrt{1 + {\xi_\text s}^{2
}} \left/ {\vrule height0.50em width0em depth0.50em} \right. \! 
 \! \Big((\eta ^{2} + \psi ^{2})\,(1 + \psi ^{2})^{2}\Big)\Bigg]{S_{22}} \\ &\,
+  \Bigg[
 - {\displaystyle \frac {3}{2}} \,{\displaystyle \frac {(3\,\eta 
^{2} - 1)\,\sqrt{1 + {\xi_\text s}^{2}}\,\mathrm{arccot}(\psi )}{
\eta ^{2} + \psi ^{2}}}  \\ &\,
\mbox{} + {\displaystyle \frac {3}{2}} \,{\displaystyle \frac {
\psi \,(5\,\eta ^{2} + 3\,\psi ^{2}\,\eta ^{2} - \psi ^{2} - 3)\,
\sqrt{1 + {\xi_\text s}^{2}}}{(\eta ^{2} + \psi ^{2})\,(1 + \psi ^{2
})^{2}}}  \Bigg]   {S_{02}}
  %%%%%%%%%%%%%%%%%%%
  .
  \end{aligned}
 \end{align}
\end{subequations} 
These source terms are fourth order polynomials in $\eta$ after having been
multiplied with the factor $(\psi^2+\eta^2)$ from eq.~\eqref{km}. Because of the
orthogonality of the Gegenbauer polynomials, $\nu_4$ is thus also a fourth order
polynomial in $\eta$. This property propagates itself through the successive
post-Newtonian orders such that a term $\phi_n$ is always an $n^\text{th}$ order
polynomial in $\eta$. Physically this is because the perturbative-like
corrections to the shape of the surface, which are in the form of a {\it finite}
sum of Legendre polynomials (see eq.~\eqref{OberGl}), give rise to a finite
number of multipole moments.

The source terms in eqs~\eqref{nu4QuellGl} contain two of the boundary 
coefficients $S_{k2}$, which have to be determined by solving
eq.~\eqref{Intformel}, i.e. requiring that the pressure vanish on the boundary.
A term containing $S_{12}$ could also be included in eqs~\eqref{nu4QuellGl},
but is found to equal zero by applying this boundary condition. Solving
eq.~\eqref{Sii} and using the abbreviation
\[ \beta:=\mathrm{arccot}(\xi_{\text s})\]
one finds
\begin{align*}
 S_{22} =&\, 
 %%%%%%%%%%
 {\displaystyle \frac {1}{2}} (1 + {\xi_{\text s}}^{2})^{(\nicefrac{3}{2})}\Big(288\,{
\xi_{\text s}}\,\beta  - 45\,\beta ^{2} + 408\,\beta \,{\xi_{\text s}}^{
3} - 54\,\beta ^{2}\,{\xi_{\text s}}^{4} + 1575\,\beta ^{2}\,{\xi _{
s}}^{8} \\&
\mbox{} - 378\,\beta ^{2}\,{\xi_{\text s}}^{2} + 1710\,\beta ^{2}\,{
\xi_{\text s}}^{6} - 3150\,{\xi_{\text s}}^{7}\,\beta  - 2370\,{\xi_{\text s}}
^{5}\,\beta  - 179\,{\xi_{\text s}}^{2} \\&
\mbox{} + 1575\,{\xi_{\text s}}^{6} + 660\,{\xi_{\text s}}^{4}\Big) \Big/\, 
\Big(3330\,
\beta \,{\xi_{\text s}}^{4} - 1965\,{\xi_{\text s}}^{3} + 732\,\beta \,{
\xi_{\text s}}^{2} \\&
\mbox{} - 357\,{\xi_{\text s}} - 5075\,{\xi_{\text s}}^{5} - 3675\,{\xi _{s
}}^{7} + 3675\,\beta \,{\xi_{\text s}}^{8} + 6300\,\beta \,{\xi_{\text s}
}^{6} + 27\,\beta \Big)
 %%%%%%%%%%%
 .
\end{align*}
The denominator of this expression is proportional to $G_4(\xi_{\text s})$ of
eq.~\eqref{Sing} and gives rise to the singularity at $\xi_4^*$, which has
already been discussed. We choose to use the remaining two non-trivial
equations extracted from eq.~\eqref{Intformel} to determine $S_{02}$ and
$\tilde{\Omega}_3$. Equivalently, we could have determined $\gamma_4$ instead
of $\tilde{\Omega}_3$. In either case, the remaining constant can be chosen
freely and does not affect the validity of the solution, but instead specifies
the Newtonian spheroid of comparison. Because the constants $S_{02}$ and
$\tilde{\Omega}_3$ are determined from a linear algebraic system of equations
involving $S_{22}$, they also contain a first order pole at the point
$\xi_4^*$.

The determination of higher orders proceeds identically. One first obtains the
Poisson-like equations by expanding eqs~(\ref{nuGl}--\ref{omegaGl}) and
extracting the coefficients of the desired order in $\varepsilon$. Next one
solves these using eq.~\eqref{IntformelGl} with $\psi$ replacing $\xi$,
integrating from 0 to $\xi_\text s$ for the interior of the star and from
$\xi_\text s$ to $\infty$ for the exterior. The metric function $\mu$ can then
be most easily computed by making use of the integral
\begin{equation}\label{muint}
 \mu-\lambda = \int_1^\eta (\mu-\lambda)_{,\eta'}\, d\eta',
\end{equation} 
where the endpoint of integration follows from
$\lim_{\eta \to 1}(\mu-\lambda)=0$ (see e.g. eqs~(22) and (23) in
\cite{Meinel02}). The integrand can be determined from the equations
\begin{align}
  \begin{split}
   \frac{1}{\varrho}(\mu-&\lambda)_{,\zeta} = (\nu+\lambda)_{,\zeta\varrho}
    +\nu_{,\varrho}\,\nu_{,\zeta} + \lambda_{,\varrho}\,\lambda_{,\zeta} -
    \mu_{,\zeta}\,(\nu+\lambda)_{,\varrho}
    \\ & - \mu_{,\varrho}\,(\nu+\lambda)_{,\zeta}
     -\frac{\tilde{\Omega}^2\varrho^2e^{2\lambda-2\nu}}{2\,{a_0}^2}
      \tilde{\omega}_{,\zeta} \tilde{\omega}_{,\varrho}
  \end{split} \\ 
  \intertext{and}
  \begin{split}
    \frac{1}{\varrho}(\mu-&\lambda)_{,\varrho} =
     \frac{1}{2}
     \left[(\nu+\lambda)_{,\varrho\varrho}-(\nu+\lambda)_{,\zeta\zeta} \right]
      +\frac{1}{2}\left(
     {\nu_{,\varrho}}^2+{\lambda_{,\varrho}}^2
       - {\nu_{,\zeta}}^2+{\lambda_{,\zeta}}^2
      \right) \\
      & -\left(\mu_{,\varrho}(\nu+\lambda)_{,\varrho}-
             \mu_{,\zeta}(\nu+\lambda)_{,\zeta}  \right)
       - \frac{\tilde{\Omega}^2\varrho^2e^{2\lambda-2\nu}}{4\,{a_0}^2}
       \left({\tilde{\omega}_{\varrho}}^2-{\tilde{\omega}_{\zeta}}^2 \right)   
   \end{split}   
 \end{align}
together with the transformation equation
\[ f_{,\eta}=a_0\xi f_{,\zeta} - 
             {a_0}^2\eta (1+\xi^2)\frac{f_{,\varrho}}{\varrho}.\] 
Because $\mu-\lambda$ is a polynomial in $\eta$ to any order of the
approximation, integrating eq.~\eqref{muint} is trivial. 
 Finally one uses the boundary condition to determine the boundary 
coefficients and $\tilde{\Omega}_{i+1}$. 

We carried out this procedure for the first four orders of the PN
approximation. The results are entirely expressible in terms of elementary
functions and in the interior of the star the metric functions are simply
polynomials with respect to $\psi^2$ and $\eta^2$. The validity of the results
was ensured by confirming that the disc limit of the expansion reduces to that
of \cite{PM}, by showing that the expressions for the gravitational mass and
angular momentum found by integrating over the interior of the star are
identical to those taken from the far field and by comparing PN-values to those
returned by highly accurate numerical calculations. 
    %%%%%%%%%%%%
   \subsection{Representative Physical Quantites}\label{Physical}
    %%%%%%%%%%%%%%%%%
    Physical quantities that are of interest in characterizing a given configuration
are its rest mass
\begin{equation}\label{Ruhemasse}
 M_0 = 2\pi Q {a_0}^3 \int_{-1}^1\int_0^{\xi_\mathrm{s}}
          \frac{e^{\lambda+2\mu}}{\sqrt{1-\tilde{v}^2}}
            \left[ \left(\psi \xi_\text B(\eta)/\xi_\mathrm{s}\right)^2+\eta^2 \right]\,
            \xi_\text B(\eta)/\xi_\mathrm{s} \; d\psi\,d\eta, 
\end{equation}
angular momentum
\begin{align}\label{Drehimpulsi}
 \begin{split}
   J =  2\pi Q {a_0}^4 c \int_{-1}^1\int_0^{\xi_\mathrm{s}}
    &\,\frac{\tilde{\Omega}(1-\gamma)(1-\tilde{\omega})e^{3\lambda-2\nu+2\mu}}
         {(1-\tilde{v}^2)^{\nicefrac{3}{2}}}[1+(\psi \xi_\text B(\eta)/\xi_\mathrm{s})^2](1-\eta^2)\\
        &\,   \left[ \left(\psi \xi_\text B(\eta)/\xi_\mathrm{s}\right)^2+\eta^2 \right]\,
            \xi_\text B(\eta)/\xi_\mathrm{s} \; d\psi\,d\eta,
 \end{split}
\end{align}
binding energy
\begin{align}\label{Energie}
 \begin{split}
  E_b =  &\, \gamma M_0\, c^2 - 2\tilde{\Omega}c/a_0\, J  \\
  &\, -4\pi {a_0}^3 \int_{-1}^1\int_0^{\xi_\mathrm{s}} P e^{\nu+\lambda+2\mu}
  \left[ \left(\psi \xi_\text B(\eta)/\xi_\mathrm{s}\right)^2+\eta^2 \right]\,
            \xi_\text B(\eta)/\xi_\mathrm{s} \; d\psi\,d\eta 
 \end{split}
\end{align}
and gravitational mass
\begin{equation}\label{Gesamtmassei}
 M = M_0 - E_b/c^2.
\end{equation}
The expression $\left[ \left(\psi \xi_\text
B(\eta)/\xi_\mathrm{s}\right)^2+\eta^2 \right]\, \xi_\text
B(\eta)/\xi_\mathrm{s} \; d\psi\,d\eta$ in the integrals comes from applying
the coordinate transformation in eq.~\eqref{OberGl} to the volume element
$(\xi^2+\eta^2)\;d\xi\,d\eta$. As an alternative to eqs~\eqref{Drehimpulsi} and
\eqref{Gesamtmassei}, one can choose to calculate the angular momentum and
gravitational mass from the far fields of $\omega$ and $\nu$ respectively and
then use eq.~\eqref{Gesamtmassei} in order to find the binding energy. The
disadvantage of the far field approach is that $\tilde{\omega}_{i+2}$ must be
known in order to find $J_i$, whereas $\tilde{\omega}_{i}$ suffices otherwise.

In addition to using the above quantities, we shall characterize configurations
by the ratio of polar to equatorial radius
\[ r_\text p/r_\text e = \xi_\text B(\eta=1)/\sqrt{1+(\xi_\text
B(\eta=0))^2},\]
the polar red shift $Z_\text p$ and ``surface potential'' $V_0$
\[
 Z_\text p =  e^{-\nu(\psi=\xi_\mathrm{s},\eta=1)}-1 \equiv e^{-V_0}-1 
  =\gamma/(1-\gamma), \]
the central pressure
\[ P_\text c = P(\psi=0,\eta=1)\]
as well as the angular velocity $\Omega$.

In appendix~\ref{tables}, tables providing information about the expansion of
these quantities can be found.
    %%%%%%%%%%%%%%%
   \subsection{Convergence and Accuracy}\label{Results}
    %%%%%%%%%%%%%%%%%%
As was mentioned in the introduction, we are lucky to have at our disposal a
highly accurate numerical code. The AKM code \cite{AKM1,AKM3} uses a
multi-domain spectral method to solve the Einstein equations for perfect fluids
in an axially symmetric, stationary spacetime for some specified equation of
state. The accuracy reached approaches machine accuracy and has thus been used
as a standard \cite{Stergioulas} to ascertain the accuracy of other numerical
codes such as Lorene/rotstar \cite{BGM}, the SF codes \cite{SF} or that of KEH
\cite{KEH1,KEH2} (see \cite{Stergioulas} for further information). Due to the
extremely high accuracy, we can use the numerically generated configurations as
though they were analytic solutions, which is what enables us to provide values
for the relative errors of physical quantities for example.

In what follows, we shall use units in which $G=c=Q=1$ holds and use the term
``Newtonian limit'' to refer to the limit in which the Newtonian and
relativistic theories agree. In this limit, $a_0$ goes to zero while $\xi_\text
s$ remains finite, thus resulting in the fact that $M \to 0$ and $e^{V_0} \to
1$. A ``Newtonian'' or ``Maclaurin'' configuration on the other hand is the term
we use to refer to the spheroidal figure that one obtains from the Newtonian
theory, i.e. from $\nu_2$ together with $\gamma_2$ and $\tilde{\Omega}_1$. 

\begin{figure}
 %%%%%%%%%%%%
 \includegraphics[width=0.9\textwidth]{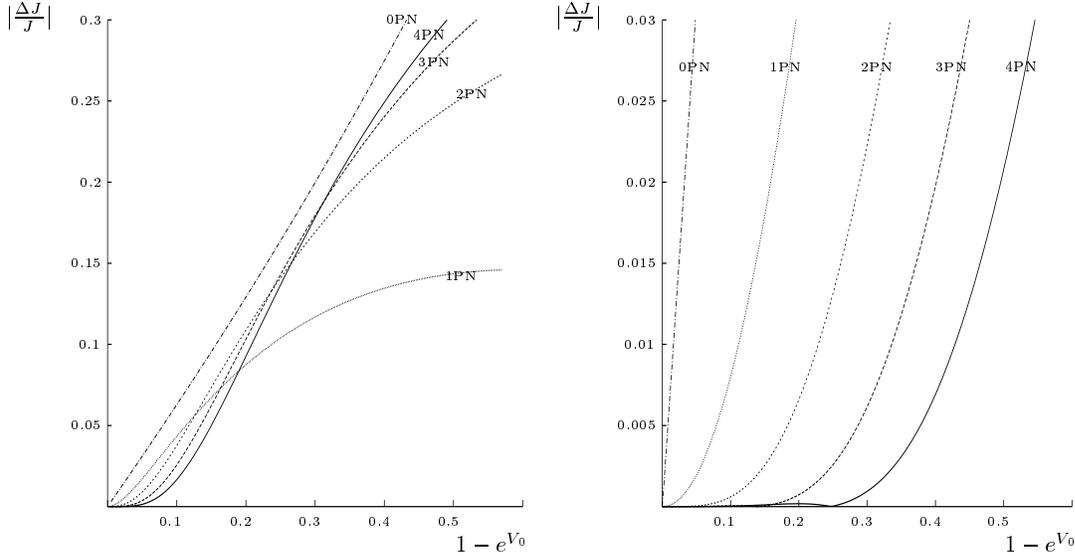}
\caption{The relative error of $J$ versus $1-e^{V_0}$ for configurations with
	 $\Omega=0.874$. On the left the PN-expansion with $\tilde{\Omega}_i=0$,
	 $i>1$, was used and on the right with $\gamma_{i+1}=0$. Here 0PN refers
	 to the Maclaurin solution, 1PN to the first order of the PN
	 approximation etc.\label{JrV}}
\end{figure}

In Table~\ref{verdelta}
\begin{table}
 \centerline{
  \begin{tabular}{l@{=\hspace{1.5mm}}llllll}
   \toprule
    \multicolumn{2}{c}{}& \multicolumn{5}{c}{Relative error for different
    choices of PN-expansion \hspace{1mm} ($i>0$)}\\
    \cmidrule(lr){3-7}
     \multicolumn{2}{l}{Numerical value}
      & $\gamma_{i+2}=0$ & $M_i=0$ & $\tilde{\Omega}_{i+1}=0$ &
     $J_i=0$ & $(r_{\text p}/r_{\text e})_i=0$  \\ 
   \midrule
     $e^{V_0}$ & $0.95$ 
      & ---    & 7.6 $\times 10^{-9}$ & ---    & 1.4 $\times 10^{-8}$ &  ---   \\
     $\Omega$ & $0.874$
      & ---    & ---    & ---    & ---    & 2.2 $\times 10^{-5}$   \\
     $M$ &  $0.004808\ldots$
      & $3.2 \times 10^{-8}$ & --- & 3.6 $\times 10^{-5}$ & 8.6 $\times 10^{-7}$
	& 2.7 $\times 10^{-5}$  \\
     $M_0$ &  $0.004936\ldots$
      & 1.0 $\times 10^{-7}$   & 4.0 $\times 10^{-7}$ &
        2.9 $\times 10^{-4}$   & 4.5 $\times 10^{-7}$  & 4.1 $\times 10^{-5}$   \\
     $P_{\mathrm c}$ &  $0.02151\ldots$
      & 1.2 $\times 10^{-6}$   &  3.6 $\times 10^{-7}$ & 7.7 $\times 10^{-4}$
	& 1.1 $\times 10^{-7}$  & 2.2 $\times 10^{-4}$ \\
     $J$  & $0.00002272\ldots$
      & 1.2 $\times 10^{-6}$   &  1.8 $\times 10^{-6}$  & 1.4 $\times 10^{-3}$
	& ---    & 3.9 $\times 10^{-4}$ \\                     
     $\frac{r_\text p}{r_\text e}$ & $0.7659\ldots$
      & 4.3 $\times 10^{-7}$ & 3.1 $\times 10^{-7}$  & 3.0 $\times 10^{-5}$
	& 1.8 $\times 10^{-7}$ & ---
    \\ \bottomrule          
  \end{tabular}}
 \caption{Relative errors for various physical quantities according to the
        $4^\text{th}$ PN order for different choices for the Newtonian
	  spheroids of comparison. The header of columns 2--6 show which
	  equation is satisfied by the respective choice for the constant 
        $\gamma_i$, $i>2$. A dash indicates that this quantity was prescribed.
	  \label{verdelta}}
\end{table}
a comparison was made of different choices for the Newtonian spheroid of
comparison. This amounts to different choices for the constants $\gamma_i$,
$i>2$. One can see in this example of a configuration near the Newtonian limit, that the
choice made can lead to differences of a few orders of magnitude for relative
errors. This surprising result can be seen, moreover, to hold over a large range of values for
the parameter $e^{V_0}$ in Figure~\ref{JrV}.
On the left hand side of this figure, the PN approximation with
$\tilde{\Omega}_i=0$, $i>1$ is depicted. The various orders react as they must
in the vicinity of the Newtonian limit $1-e^{V_0}=0$: each new order brings
about a noticeable improvement. As one moves away from this limit however, the
curves cross each other and it turns out that higher orders render a
worse approximation than lower ones. The right hand side of the figure tells a very
different story. Here the PN approximation with $\gamma_{i+1}=0$, $i>1$ is
depicted. Each additional term in the PN approximation brings about a
marked improvement in accuracy and, moreover, the relative error is more
than an order of magnitude lower than on the left hand side.

\begin{figure}
 %%%%%%%%%%%%%
 \includegraphics[width=0.9\textwidth]{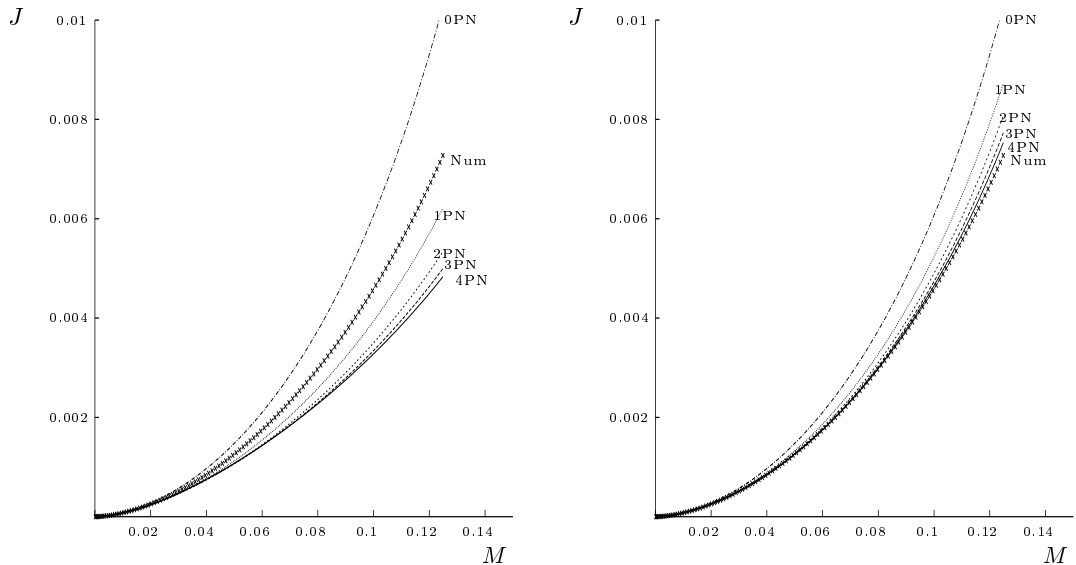}
\caption{A plot of $J$ over $M$ for configurations with
	 $\Omega=0.874$. On the left the PN-expansion with $\tilde{\Omega}_i=0$,
	 $i>1$, was used and on the right with $\gamma_{i+1}=0$. Num refers
	 to the numerical solution, 0PN
	 to the Maclaurin solution, 1PN to the first order of the PN
	 approximation etc.\label{JuM}}
\end{figure}

Imagine for a moment that one had calculated the PN approximation presented
here without being in possession of numerical values. Furthermore, let us
imagine that one had decided from the outset to prescribe $\tilde{\Omega}_i=0$,
$i>1$. Then one would have been able to produce the plot on the left hand side
of Figure~\ref{JuM} without the numerical curve. It would have been natural to
suppose that the PN series converges toward the correct solution and that
the fourth order of the PN approximation almost provides the correct value for
$J$ even up to values for $M$ of 0.12.  Had one chosen $\gamma_{i+1}=0$ instead,
then one would have produced the right hand side of Figure~\ref{JuM} without the
numerical curve and come to
the same conclusions regarding the convergence of the PN approximation. In that
case, however, one would have been correct. 

Although curves depicting relative errors of other physical quantities may look
quite different from those for $J$ shown in Figure~\ref{JrV}, they also have
many important aspects in common. The choice $\gamma_{i+1}=0$, $i>1$, leads to
much smaller relative errors than for $\tilde{\Omega}_{i}=0$ and one tends to
find improvement with increasing order even far away from the Newtonian limit.
These properties hold for a wide range of $\Omega$ values and the relative
errors tend to decrease with decreasing angular velocity so long as one does not
come too close to a singularity in the parameter space. 

\begin{figure}
 %%%%%%%%%%
 \centerline{\includegraphics[width=0.9\textwidth]{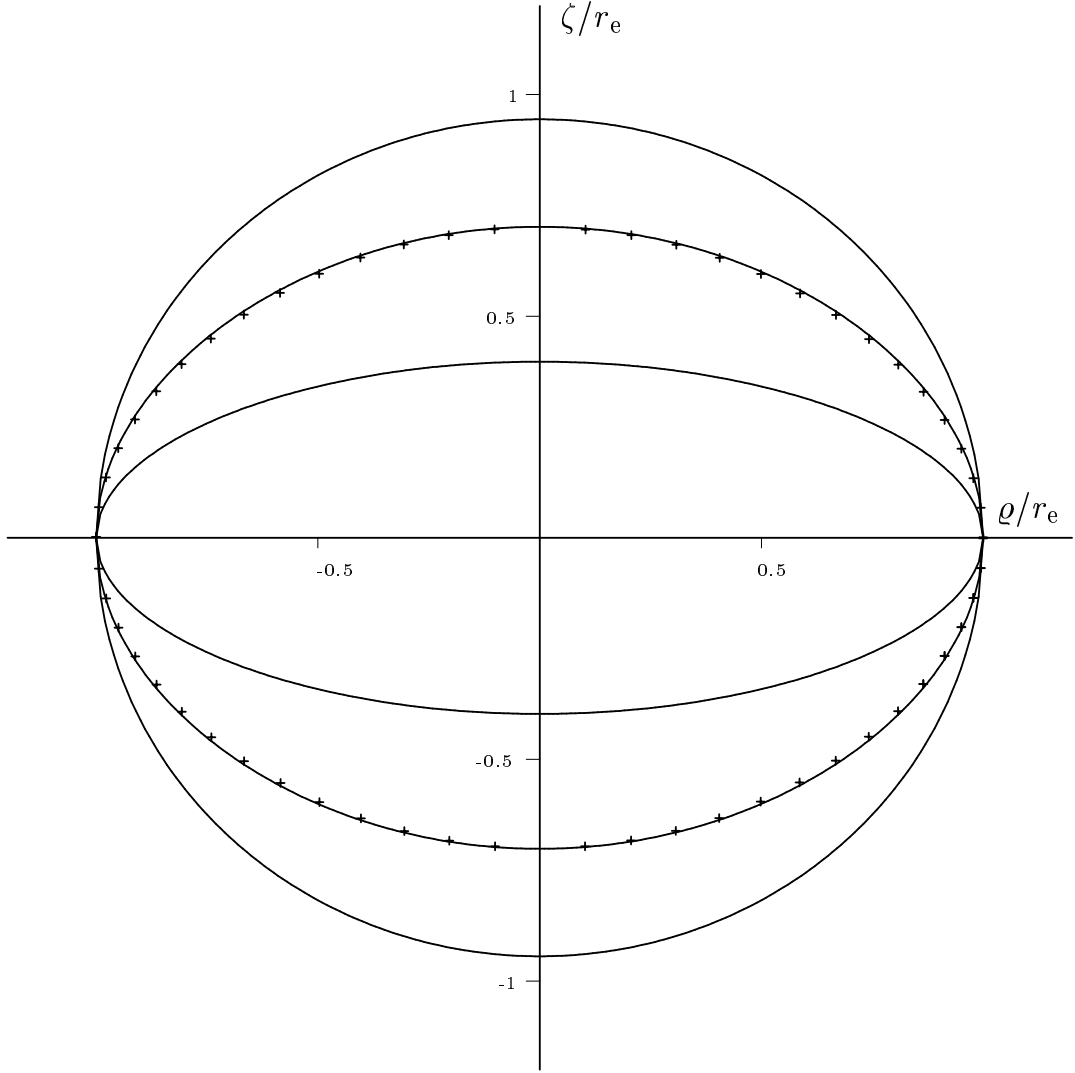}}
\caption{The meridional cross-section of a highly relativistic configuration. For fixed values of
$\xi_\text s$ and $a_0$, the outermost curve represents the Maclaurin spheroid, the innermost was
calculated using the first order approximation and the curve between these using the fourth order
approximation. The crosses are numerical values.\label{Nozawa}}
\end{figure}

One well known technique for improving on the PN approximation is the use of
the Pad{\'e} approximant, which approximates a truncated series by a quotient of
two polynomials and is discussed with reference to the PN approximation in
\cite{Damour}. For the disc limit of the solution considered here, it has been shown in
\cite{PM} that the Pad{\'e} approximant provides a far better approximation of
the analytic solution given in \cite{NM95,NM03} than the PN
approximation itself. In the case of the Maclaurin spheroids this turns out to be
true as well, especially for $\Omega<0.8$. We see in Table~\ref{Om3e7} 
\begin{table}
 \centerline{\begin{tabular}{llllllll} \toprule
 & 0PN & 1PN & 2PN & 3PN & 4PN & Pad{\'e} & AKM \\
  \midrule
  $e^{V_0}$ & 0.7 & 0.7 & 0.7 & 0.7 & 0.7 & 0.7 & 0.7  \\
   $\Omega$  & 0.3 & 0.3 & 0.3 & 0.3 & 0.3 & 0.3 & 0.3  \\
   $M\; (\times 10^{-2})$ &  
       7.94 & 6.17 & 6.244 & 6.2490 & 6.25013 & 6.25055 & 6.25070 \\
   $M_0\; (\times 10^{-2})$ &
       7.94 & 7.60 & 7.589 & 7.5864 & 7.58590 & 7.58554 & 7.58553 \\
   $P_\text c\; (\times 10^{-1})$ &
       1.47 & 2.14 & 2.434 & 2.5627 & 2.61841 & 2.66099 & 2.66064 \\
   $J\; (\times 10^{-4})$ &
       6.90 & 6.34 & 6.141 & 6.0744 & 6.05313 & 6.04321 & 6.04352 \\
   $\frac{r_\text p}{r_\text e}\; (\times 10^{-1})$ &
       9.73 & 9.79 & 9.801 & 9.8041 & 9.80538 & 9.80606 & 9.80628 
 \\ \bottomrule
 \end{tabular}}
 \caption{Values of various physical quantities according to different orders of
 the PN approximation. 0PN stands for the Maclaurin solution, 1PN for the first
 PN approximation etc. The PN approximation with $\gamma_i=0$, $i>2$ was used
 and the Pad{\'e} approximant was applied to the fourth order solution.  \label{Om3e7}}
\end{table}
how well the Pad{\'e} approximant with a polynomial of sixth order in the
numerator and second order in the denominator converges to the correct solution. Most likely
this technique would be even more effective when applied to a somewhat higher
order of the approximation. Even up to the fourth order, the PN approximation
turns out to be roughly comparable to older numerical codes even for highly
relativistic configurations. An impressive illustration of its applicability in
such highly relativistic regimes can be found in Fig.~\ref{Nozawa}.
In this figure, the meridional cross section of a configuration with a central pressure of 1 and
a radius ratio of 0.7 is depicted. One can see that the surface predicted by the fourth order PN
approximation is almost indistinguishable from the numerical values.

For a more detailed comparison with numerical values and a more complete
account regarding the derivation of the iterative scheme and the singularities
in parameter space, the reader is referred to \cite{P}
    %%%%%%%%%%%%%%%%%%    
  
  \section{Conclusion}
   %%%%%%%%%%%%%%%%
   In this paper, an iterative procedure to enable the explicit calculation of any
order of the PN approximation of the Maclaurin spheroids was devised. This was
made possible by introducing coordinates tailored to the unknown surface of the
star, by requiring that this surface's representation be a terminating sum and
by realizing that eq.~\eqref{IntformelGl} can be used in the new coordinates
without alteration. The PN expansion was carried out explicitly to the fourth
order and the resulting expressions contained only elementary functions.

It was proved that the $n^{\text{th}}$ PN approximation has a first order pole
at $\xi_{2n+2}^*$, the onset of the $n^{\text{th}}$ axisymmetric, harmonic mode
of secular instability. The radius of convergence of the series becomes zero at
these points, thereby dividing the $\xi_\text s$-$\varepsilon$ parameter space
into rectangles with ``impermeable'' walls that accumulate about (but not at)
the line $\xi_\text s=0$. Since the PN approximation appears to converge even in the highly
relativistic regime, it seems likely that no quasi-stationary, axisymmetric
sequence of solutions leads from an extended, three dimensional configuration
to the disc limit -- all such configurations would have to pass through an
infinite number of such impermeable walls.

The convergence of the PN approximation was shown to depend strongly on the
choice of the Newtonian configuration of comparison. A poor choice can render
the approximation useless in the relativistic regime, but a good one was shown to
converge quite well, especially when aided by the Pad{\'e} approximant. These
results can be taken as a word of warning, reminding the researcher that the PN
approximation can be very sensitive to alterations that may have no direct
physical consequences. On the other hand, they also demonstrate that in the best
of circumstances, the PN expansion can yield a very good approximation to highly
relativistic configurations, well beyond its guaranteed region of validity.
   %%%%%%%%%%%%%%

 \begin{appendix}
  %%%%%%%%%%%%%%%%%
  \section{The functions $g_l^m(\psi)$ and $h_l^m(\psi)$}\label{gh}
 \begin{align*}
  g_0^2 =& \, 1 \\
  g_2^2 =& \, 
  %%%%
  - {\displaystyle \frac {3\,\psi ^{2}}{2}}  - {\displaystyle 
\frac {1}{2}} 
  %%%%%
   \\ 
  g_4^2 =& \, 
  %%%%%%%%
  {\displaystyle \frac {35}{8}} \,\psi ^{4} + {\displaystyle 
\frac {15}{4}} \,\psi ^{2} + {\displaystyle \frac {3}{8}} 
  %%%%%%%
   \\ 
  g_6^2 =& \, 
  %%%%%%%%%
   - {\displaystyle \frac {231}{16}} \,\psi ^{6} - {\displaystyle 
\frac {315}{16}} \,\psi ^{4} - {\displaystyle \frac {105}{16}} \,
\psi ^{2} - {\displaystyle \frac {5}{16}} 
  %%%%%%%%%
   \\  \\
  h_0^2 =& \, 
  %%%%%%%%%%
  \mathrm{arccot}(\psi)
  %%%%%%%%
   \\ 
  h_2^2 =& \, 
  %%%%%%%%%
  {\displaystyle \frac {3\,\psi }{2}}  + \left( - {\displaystyle \frac {
3\,\psi ^{2}}{2}}  - {\displaystyle \frac {1}{2}} \right)\,\mathrm{
arccot}(\psi )
  %%%%%%%%%%
   \\ 
  h_4^2 =& \,
  %%%%%%%
   - {\displaystyle \frac {35\,\psi ^{3}}{8}}  - {\displaystyle 
\frac {55\,\psi }{24}}  + \left({\displaystyle \frac {35}{8}} \,\psi 
^{4} + {\displaystyle \frac {15}{4}} \,\psi ^{2} + 
{\displaystyle \frac {3}{8}} \right)\,\mathrm{arccot}(\psi )
  %%%%%%%%
  \\ 
  h_6^2 =& \, 
  %%%%%%%%%%
  {\displaystyle \frac {231\,\psi ^{5}}{16}}  + {\displaystyle 
\frac {119\,\psi ^{3}}{8}}  + {\displaystyle \frac {231\,\psi }{
80}} \\&
+ \left( - {\displaystyle \frac {231}{16}} \,\psi ^{6} - 
{\displaystyle \frac {315}{16}} \,\psi ^{4} - {\displaystyle 
\frac {105}{16}} \,\psi ^{2} - {\displaystyle \frac {5}{16}} \right)\,
\mathrm{arccot}(\psi )
  %%%%%%%%%%%
   \\ \\
  g_0^3 =& \, 1 \\
  g_2^3 =& \, 
  %%%%%%%
  - 4\,\psi ^{2} - 1
  %%%%%%%
    \\
  g_4^3 =& \, 
  %%%%%%%%%
  16\,\psi ^{4} + 12\,\psi ^{2} + 1
  %%%%%%%%%
    \\
  g_6^3 =& \, 
  %%%%%%%%%
  - 64\,\psi ^{6} - 80\,\psi ^{4} - 24\,\psi ^{2} - 1
  %%%%%%%%%%
  \\ \\
  h_0^3 =& \, 
  %%%%%%%%%%%%%%%%
  1 - {\displaystyle \frac {\psi }{\sqrt{1 + \psi ^{2}}}}
  %%%%%%%%%%%%
   \\
  h_2^3 =& \, 
  %%%%%%%%%%
  - {\displaystyle \frac {4\,\psi ^{2}}{3}}  - {\displaystyle 
\frac {1}{3}}  + {\displaystyle \frac {{\displaystyle \frac {4}{3
}} \,\psi ^{3} + \psi }{\sqrt{\psi ^{2} + 1}}}
  %%%%%%%%%
   \\
  h_4^3 =& \, 
  %%%%%%%%%%%
  {\displaystyle \frac {16\,\psi ^{4}}{5}}  + {\displaystyle 
\frac {12\,\psi ^{2}}{5}}  + {\displaystyle \frac {1}{5}}  + 
{\displaystyle \frac { - {\displaystyle \frac {16}{5}} \,\psi ^{5
} - 4\,\psi ^{3} - \psi }{\sqrt{\psi ^{2} + 1}}}
  %%%%%%%%%%%
    \\
  h_6^3 =& \, 
  %%%%%%%%
  - {\displaystyle \frac {64\,\psi ^{6}}{7}}  - {\displaystyle 
\frac {80\,\psi ^{4}}{7}}  - {\displaystyle \frac {24\,\psi ^{2}
}{7}}  - {\displaystyle \frac {1}{7}}  + {\displaystyle \frac {
{\displaystyle \frac {64}{7}} \,\psi ^{7} + 16\,\psi ^{5} + 8\,
\psi ^{3} + \psi }{\sqrt{\psi ^{2} + 1}}}
  %%%%%%%%%
   \\ \\
  g_0^4 =& \, 1 \\ 
  g_2^4 =& \,
   %%%%%%%
   - {\displaystyle \frac {15\,\psi ^{2}}{2}}  - {\displaystyle 
\frac {3}{2}} 
   %%%%%%%
   \\
  g_4^4 =& \, 
  %%%%%%%%%
  {\displaystyle \frac {315}{8}} \,\psi ^{4} + {\displaystyle 
\frac {105}{4}} \,\psi ^{2} + {\displaystyle \frac {15}{8}} 
  %%%%%%%%%
    \\
  g_6^4 =& \, 
  %%%%%%%
  - {\displaystyle \frac {3003}{16}} \,\psi ^{6} - {\displaystyle 
\frac {3465}{16}} \,\psi ^{4} - {\displaystyle \frac {945}{16}} 
\,\psi ^{2} - {\displaystyle \frac {35}{16}}
  %%%%%%%%%
    \\ \\
  h_0^4 =& \, 
  %%%%%%%%
  {\displaystyle \frac {1}{2}} \,\mathrm{arccot}(\psi ) - 
{\displaystyle \frac {\psi }{2\,(\psi ^{2} + 1)}} 
  %%%%%%%%
    \\
  h_2^4 =& \, 
  %%%%%%%%%%
  \left( - {\displaystyle \frac {5\,\psi ^{2}}{8}}  - {\displaystyle 
\frac {1}{8}} \right)\,\mathrm{arccot}(\psi ) + {\displaystyle \frac {(
15\,\psi ^{2} + 13)\,\psi }{24\,(\psi ^{2} + 1)}} 
  %%%%%%%%
    \\
  h_4^4 =& \, 
  %%%%%%%%%
  \left({\displaystyle \frac {21}{16}} \,\psi ^{4} + {\displaystyle 
\frac {7}{8}} \,\psi ^{2} + {\displaystyle \frac {1}{16}} \right)\,
\mathrm{arccot}(\psi ) - {\displaystyle \frac {(315\,\psi ^{4} + 
420\,\psi ^{2} + 113)\,\psi }{240\,(\psi ^{2} + 1)}} 
  %%%%%%%%%
    \\
  h_6^4 =& \, 
  %%%%%%%%
  \left( - {\displaystyle \frac {429}{128}} \,\psi ^{6} - 
{\displaystyle \frac {495}{128}} \,\psi ^{4} - {\displaystyle 
\frac {135}{128}} \,\psi ^{2} - {\displaystyle \frac {5}{128}} \right)
\,\mathrm{arccot}(\psi ) \\ &
\mbox{} + {\displaystyle \frac {(15015\,\psi ^{6} + 27335\,\psi 
^{4} + 14273\,\psi ^{2} + 1873)\,\psi }{4480\,(\psi ^{2} + 1)}} 
  %%%%%%%%  
\end{align*}
 
\section{Tables of Various Physical Quantities}\label{tables}
 
This appendix contains tables with the numerical values for the post-Newtonian
coefficients of the quantities introduced in section~\ref{Physical} for various
values of $\xi_\text s$. In all the tables, we have chosen $\gamma_i=0$, $i>2$,
whence we find
\[
\gamma = \gamma_2 \varepsilon^2 =
\frac{3}{4}\sqrt{1+{\xi_\mathrm{s}}^2}\left(\ar(\xi_\mathrm{s})(1+{\xi_\mathrm{s}}^2)-\xi_\mathrm{s}
\right)\,\varepsilon^2.
\]
Taking into account
\[ {a_0}^2 = \frac{3 c^2}{8\pi\,Q
\xi_\mathrm{s}\sqrt{1+\xi_\mathrm{s}^{\,2}}}\, \varepsilon^2, \]
we would find, for example, the following values for the second PN approximation
of configuration with $\xi_\text s=0.5$, $\varepsilon=0.7$:
\begin{align*}
 a_0 &\approx\, 0.32346  \\
 e^{V_0} &=\, 1-\gamma \approx\, 0.63681  \\
 \Omega  &= \tilde{\Omega}/a_0 \approx \frac{1}{a_0}
             \left( 0.54175\,\varepsilon + 0.26458\,\varepsilon^3 -
		      0.32261\,\varepsilon^5\right)
		  \approx \, 1.285 \\
 M &\approx 
             \left(2.6180  - 1.3693\, \varepsilon^2 +
		      1.7071\, \varepsilon^4\right)\, {a_0}^3
		  \approx \, 0.07976 \\
 M_0 &\approx \left(2.6180  - 0.58922\, \varepsilon^2 +
		      0.99562\, \varepsilon^4\right)\, {a_0}^3
		  \approx \, 0.08692 \\
 P_\text c &\approx \left(0.87655  - 0.29553\, \varepsilon^2 -
		      0.21693\, \varepsilon^4\right) {a_0}^2
		  \approx \, 0.07111 \\
 J &\approx \left(1.5346  + 1.0605\, \varepsilon^2 +
		      0.24311\, \varepsilon^4\right)\, {a_0}^5
		  \approx \, 0.007480 \\
 \frac{r_\text p}{r_\text e} 
    &\approx \left(0.44721  - 0.55974\, \varepsilon^2 -
		      0.23039\, \varepsilon^4\right)
		  \approx \, 0.2283.
\end{align*}

\renewcommand{\arraystretch}{0.90}

\begin{table}
  \centerline{
   \begin{tabular}{llllll} \toprule
     & \multicolumn{5}{c}
      {Coefficients of $\varepsilon^i$ for the dimensionless angular velocity
	$\tilde{\Omega}$}\\
     \cmidrule(lr){2-6}
    $\xi_\mathrm{s}$ & $\varepsilon^1$ & $\varepsilon^3$ & $\varepsilon^5$ 
           & $\varepsilon^7$ & $\varepsilon^9$ \\ \midrule 
    %%%%%%%%%%%%%%%
    0.00         	&	1.0854019 	&	-0.95903173	&	-0.21316642	&	-0.09051296	&	-0.04918981	\\
0.01      	&	1.0716302 	&	-0.92995651	&	-0.19734833	&	-0.11039829	&	-0.03104705	\\
0.02      	&	1.0579561 	&	-0.90160729	&	-0.18278329	&	-0.12768605	&	-0.01297664	\\
0.03      	&	1.0443818 	&	-0.8741343 	&	-0.16911244	&	-0.14231975	&	0.00475476	\\
0.04      	&	1.0309094 	&	-0.84772939	&	-0.15613789	&	-0.15411481	&	0.02219393	\\
0.05      	&	1.0175410 	&	-0.82264177	&	-0.14417187	&	-0.16467618	&	0.01858310	\\
0.06      	&	1.0042785 	&	-0.7992021 	&	-0.13496336	&	-0.19414756	&	-1.0164212 	\\
0.07      	&	0.99112387	&	-0.77786047	&	-0.13435142	&	-0.44265195	&	24.323369  	\\
0.08      	&	0.97807889	&	-0.75924895	&	-0.16091713	&	-5.9366901 	&	413.17220   	\\
0.09      	&	0.96514535	&	-0.74428879	&	-0.28117178	&	12.104546  	&	-22.734741  	\\
0.10       	&	0.95232494	&	-0.73438506	&	-0.8296993 	&	31.560268  	&	-462.39599   	\\
0.11      	&	0.93961928	&	-0.73180435	&	-8.5365954 	&	809.03790   	&	-58587.792     	\\
0.12      	&	0.92702988	&	-0.74047382	&	5.2676596 	&	10.231313  	&	2190.3915    	\\
0.13      	&	0.91455822	&	-0.76787434	&	4.9354647 	&	-55.887234  	&	1224.6781    	\\
0.14      	&	0.90220565	&	-0.83029178	&	7.3381300 	&	-134.37904   	&	3062.2576    	\\
0.15      	&	0.88997346	&	-0.97129085	&	14.854009  	&	-437.85365   	&	15041.748     	\\
0.16      	&	0.87786287	&	-1.3603363 	&	49.619624  	&	-3139.3425    	&	234545.37      	\\
0.17      	&	0.865875  	&	-3.9363526 	&	1171.2187    	&	-606631.48      	&	3.87526090
$\times 10^8$	\\
0.18      	&	0.85401088	&	1.9317681 	&	-11.780334  	&	2884.4373    	&	-326822.46      	\\
0.19      	&	0.84227147	&	0.57651868	&	13.454904  	&	127.48863   	&	100.77384   	\\
0.20       	&	0.83065764	&	0.27773007	&	6.1769373 	&	69.330710  	&	524.99073   	\\
0.30       	&	0.72155838	&	0.09314481	&	-0.2608214 	&	1.6431116 	&	5.7294717 	\\
0.40       	&	0.62537562	&	0.1993343 	&	-0.43302451	&	0.15006752	&	1.3582894 	\\
0.50       	&	0.54174791	&	0.26458306	&	-0.32260971	&	-0.27140215	&	0.58262038	\\
0.60       	&	0.46979598	&	0.28819495	&	-0.17835186	&	-0.34929691	&	0.09017830	\\
0.70       	&	0.40833633	&	0.28520112	&	-0.06646413	&	-0.28550798	&	-0.13370899	\\
0.80       	&	0.35606582	&	0.26835131	&	0.00483724	&	-0.19391875	&	-0.17918978	\\
0.90       	&	0.3116946 	&	0.24576778	&	0.04504743	&	-0.11763146	&	-0.15297576	\\
1.00         	&	0.27402709	&	0.22194776	&	0.06517545	&	-0.06422083	&	-0.11177801	\\
1.10 	&	0.2420016 	&	0.19912131	&	0.07339547	&	-0.02967889	&	-0.07575039	\\
1.20 	&	0.21470198	&	0.17824581	&	0.07493751	&	-0.00829155	&	-0.04899455	\\
1.30 	&	0.19135249	&	0.15961339	&	0.07292802	&	0.00454162	&	-0.0304128 	\\
1.40 	&	0.17130433	&	0.14319157	&	0.06916914	&	0.01198574	&	-0.01789323	\\
1.50 	&	0.15401892	&	0.1288049 	&	0.06467418	&	0.01608587	&	-0.00957769	\\
1.60 	&	0.13905099	&	0.11622792	&	0.05999884	&	0.01812893	&	-0.00409702	\\
1.70 	&	0.12603306	&	0.10523057	&	0.05543697	&	0.01891680	&	-0.00050821	\\
1.80 	&	0.11466173	&	0.09559856	&	0.05113403	&	0.01894809	&	0.00182048	\\
1.90 	&	0.10468625	&	0.08714120	&	0.04715237	&	0.01853317	&	0.00330817	\\
2.00         	&	0.09589882	&	0.07969281	&	0.04350861	&	0.0178655 	&	0.00423273	\\
    %%%%%%%%%%%%%%%%%
   \bottomrule
   \end{tabular}}
   \caption{Expansion coefficients of $\tilde{\Omega}$
         for various values of $\xi_\text s$ with $\gamma_i=0$, $i>2$.
	   \label{Oanhang}}
 \end{table}

 \begin{table}
  \centerline{
   \begin{tabular}{llllll}\toprule
     & \multicolumn{5}{c}
     {Coefficients of $\varepsilon^i$ for the gravitational mass $M$}\\
    \cmidrule(lr){2-6}
    $\xi_\mathrm{s}$ & $\varepsilon^0$ & $\varepsilon^2$ & $\varepsilon^4$ 
           & $\varepsilon^6$ & $\varepsilon^8$ \\ \midrule 
    %%%%%%%%%%%%%%%%
    0.00         	&	0	&	0	&	0	&	0	&	0	\\
0.01      	&	0.04189209	&	0.04246214	&	0.04840595	&	0.05310172	&	0.06138033	\\
0.02      	&	0.08380931	&	0.08115227	&	0.09502374	&	0.09931087	&	0.1165261 	\\
0.03      	&	0.1257768 	&	0.11627705	&	0.14034491	&	0.13816091	&	0.16779538	\\
0.04      	&	0.16781969	&	0.14806521	&	0.18487986	&	0.1690188 	&	0.2181801 	\\
0.05      	&	0.20996311	&	0.17677919	&	0.22928743	&	0.19157084	&	0.27677559	\\
0.06      	&	0.25223219	&	0.2027326 	&	0.27473656	&	0.21060612	&	0.5979699 	\\
0.07      	&	0.29465207	&	0.22631723	&	0.32398279	&	0.28537561	&	-6.0060957 	\\
0.08      	&	0.33724788	&	0.24804689	&	0.38498653	&	2.0152694 	&	-124.06068   	\\
0.09      	&	0.38004475	&	0.26863219	&	0.48597184	&	-3.9985889 	&	-0.15355203	\\
0.10       	&	0.42306781	&	0.28911548	&	0.77113254	&	-11.621492  	&	155.23679   	\\
0.11      	&	0.4663422 	&	0.31113215	&	4.1859186 	&	-339.22981   	&	24417.212     	\\
0.12      	&	0.50989305	&	0.33746264	&	-1.8588230 	&	-10.933727  	&	-1022.9939    	\\
0.13      	&	0.5537455 	&	0.37333934	&	-1.7708758 	&	21.555107  	&	-531.17370   	\\
0.14      	&	0.59792467	&	0.43007081	&	-3.0839441 	&	61.463611  	&	-1420.9588    	\\
0.15      	&	0.6424557 	&	0.5377904 	&	-7.4131614 	&	226.89235   	&	-7777.7392    	\\
0.16      	&	0.68736372	&	0.81351169	&	-28.767430  	&	1830.3581    	&	-136275.58      	\\
0.17      	&	0.73267386	&	2.5952925 	&	-770.75767   	&	398608.22      	&	-2.54467540
$\times 10^8$	\\
0.18      	&	0.77841126	&	-1.4529533 	&	11.397074  	&	-2110.8820    	&	244860.84      	\\
0.19      	&	0.82460105	&	-0.51954369	&	-10.093076  	&	-89.164129  	&	226.04238   	\\
0.20       	&	0.87126836	&	-0.3180264 	&	-4.8968261 	&	-57.161748  	&	-380.38457   	\\
0.30       	&	1.3697344 	&	-0.37666528	&	0.81904278	&	-2.7961160 	&	-9.8763842 	\\
0.40       	&	1.9435987 	&	-0.82904344	&	1.5102380 	&	-0.6053959 	&	-3.4289311 	\\
0.50       	&	2.6179939 	&	-1.3693167 	&	1.7070560 	&	0.53885692	&	-2.1926160 	\\
0.60       	&	3.4180528 	&	-1.9472924 	&	1.6481885 	&	1.2005744 	&	-0.98288389	\\
0.70       	&	4.3689082 	&	-2.5527128 	&	1.4829837 	&	1.4507518 	&	-0.00014983	\\
0.80       	&	5.4956927 	&	-3.1933102 	&	1.3002647 	&	1.4514458 	&	0.57039853	\\
0.90       	&	6.8235392 	&	-3.8834294 	&	1.1398464 	&	1.3414215 	&	0.81909493	\\
1.00         	&	8.3775804 	&	-4.6386450 	&	1.0140861 	&	1.1999932 	&	0.88499186	\\
1.10 	&	10.182949  	&	-5.4737714 	&	0.92337032	&	1.0630189 	&	0.86508412	\\
1.20 	&	12.264778  	&	-6.4024064 	&	0.86407489	&	0.94327903	&	0.81216639	\\
1.30 	&	14.648199  	&	-7.4370306 	&	0.83193824	&	0.84322071	&	0.75142342	\\
1.40 	&	17.358347  	&	-8.5892485 	&	0.82324479	&	0.76132333	&	0.69360857	\\
1.50 	&	20.420352  	&	-9.8700213 	&	0.83510223	&	0.69491426	&	0.64259313	\\
1.60 	&	23.859349  	&	-11.289852  	&	0.86539261	&	0.64129035	&	0.59918663	\\
1.70 	&	27.700470  	&	-12.858921  	&	0.91263339	&	0.59809715	&	0.56295176	\\
1.80 	&	31.968847  	&	-14.587185  	&	0.97583498	&	0.56340574	&	0.53302851	\\
1.90 	&	36.689613  	&	-16.484440  	&	1.0543810 	&	0.53567923	&	0.50848981	\\
2.00         	&	41.887902  	&	-18.560374  	&	1.1479349 	&	0.51370787	&	0.48848149	\\
    %%%%%%%%%%%%%%%%
   \bottomrule
   \end{tabular}}
   \caption{Expansion coefficients of $\frac{M}{{a_0}^3\,Q}$
         for various values of $\xi_\text s$ with $\gamma_i=0$, $i>2$.}
 \end{table}
 
 \begin{table}
  \centerline{
   \begin{tabular}{llllll} \toprule
     & \multicolumn{5}{c}
     {Coefficients of $\varepsilon^i$ for the rest mass $M_0$}\\
    \cmidrule(lr){2-6}
    $\xi_\mathrm{s}$ & $\varepsilon^0$ & $\varepsilon^2$ & $\varepsilon^4$ 
           & $\varepsilon^6$ & $\varepsilon^8$ \\ \midrule 
    %%%%%%%%%%%%%%%%%%%
    0.00         	&	0	&	0	&	0	&	0	&	0	\\
0.01      	&	0.04189209	&	0.05245596	&	0.06452339	&	0.07673604	&	0.09064350	\\
0.02      	&	0.08380931	&	0.10138291	&	0.12467777	&	0.14552079	&	0.17045889	\\
0.03      	&	0.1257768 	&	0.14697999	&	0.1810503 	&	0.20649775	&	0.24117379	\\
0.04      	&	0.16781969	&	0.18946928	&	0.23424172	&	0.25967447	&	0.30478958	\\
0.05      	&	0.20996311	&	0.22910739	&	0.28499562	&	0.30546133	&	0.36899426	\\
0.06      	&	0.25223219	&	0.26620288	&	0.33455944	&	0.34955489	&	0.68695096	\\
0.07      	&	0.29465207	&	0.30114322	&	0.38575957	&	0.4525081 	&	-5.9229587 	\\
0.08      	&	0.33724788	&	0.3344386 	&	0.44661698	&	2.2159335 	&	-123.76573   	\\
0.09      	&	0.38004475	&	0.36679666	&	0.5454014 	&	-3.7541975 	&	-0.6779804 	\\
0.10       	&	0.42306781	&	0.39925742	&	0.82632859	&	-11.306658  	&	154.00435   	\\
0.11      	&	0.4663422 	&	0.43345453	&	4.2348294 	&	-338.55795   	&	24389.392     	\\
0.12      	&	0.50989305	&	0.47216724	&	-1.8183462 	&	-10.679058  	&	-1027.1961    	\\
0.13      	&	0.5537455 	&	0.52062728	&	-1.7412446 	&	21.991097  	&	-535.19617   	\\
0.14      	&	0.59792467	&	0.59014308	&	-3.0682419 	&	62.144578  	&	-1430.4764    	\\
0.15      	&	0.6424557 	&	0.71084828	&	-7.4164107 	&	228.15983   	&	-7811.5324    	\\
0.16      	&	0.68736372	&	0.99975724	&	-28.802435  	&	1834.1087    	&	-136523.76      	\\
0.17      	&	0.73267386	&	2.7949289 	&	-770.92714   	&	398680.63      	&	-2.54506740
$\times 10^8$	\\
0.18      	&	0.77841126	&	-1.2397211 	&	11.483888  	&	-2103.2027    	&	244823.57      	\\
0.19      	&	0.82460105	&	-0.29250893	&	-10.086248  	&	-86.857421  	&	270.08015   	\\
0.20       	&	0.87126836	&	-0.07698001	&	-4.9219300 	&	-55.908647  	&	-364.07926   	\\
0.30       	&	1.3697344 	&	0.01672239	&	0.56758844	&	-2.1242466 	&	-9.5006993 	\\
0.40       	&	1.9435987 	&	-0.25789658	&	1.0217665 	&	0.07622924	&	-2.9245661 	\\
0.50       	&	2.6179939 	&	-0.58921661	&	0.99559791	&	1.0782019 	&	-1.3199037 	\\
0.60       	&	3.4180528 	&	-0.92033156	&	0.74161054	&	1.5062120 	&	0.04353243	\\
0.70       	&	4.3689082 	&	-1.2337985 	&	0.40781773	&	1.5054690 	&	0.95554101	\\
0.80       	&	5.4956927 	&	-1.5299374 	&	0.07578544	&	1.2771041 	&	1.3384899 	\\
0.90       	&	6.8235392 	&	-1.8155767 	&	-0.22248284	&	0.97101036	&	1.3676277 	\\
1.00         	&	8.3775804 	&	-2.0987311 	&	-0.48106395	&	0.66449818	&	1.2237982 	\\
1.10 	&	10.182949  	&	-2.3866409 	&	-0.70436865	&	0.38738975	&	1.0182639 	\\
1.20 	&	12.264778  	&	-2.6853269 	&	-0.89947761	&	0.14629252	&	0.80533928	\\
1.30 	&	14.648199  	&	-2.9996963 	&	-1.0731453 	&	-0.06144076	&	0.60709587	\\
1.40 	&	17.358347  	&	-3.3337847 	&	-1.2309272 	&	-0.24120271	&	0.43012826	\\
1.50 	&	20.420352  	&	-3.6909893 	&	-1.3771251 	&	-0.39850746	&	0.27444199	\\
1.60 	&	23.859349  	&	-4.0742533 	&	-1.5149872 	&	-0.53811594	&	0.13764981	\\
1.70 	&	27.700470  	&	-4.4862014 	&	-1.6469456 	&	-0.66387001	&	0.01679751	\\
1.80 	&	31.968847  	&	-4.9292361 	&	-1.7748240 	&	-0.7787784 	&	-0.09091482	\\
1.90 	&	36.689613  	&	-5.4056047 	&	-1.8999983 	&	-0.88516836	&	-0.18789822	\\
2.00         	&	41.887902  	&	-5.9174456 	&	-2.0235176 	&	-0.98483424	&	-0.27614163	\\
    %%%%%%%%%%%%%%%
   \bottomrule
   \end{tabular}}
   \caption{Expansion coefficients of $\frac{M_0}{{a_0}^3\,Q}$
         for various values of $\xi_\text s$ with $\gamma_i=0$, $i>2$.}
 \end{table}

\begin{table}
  \centerline{
   \begin{tabular}{llllll} \toprule
     & \multicolumn{5}{c}
     {Coefficients of $\varepsilon^i$ for the central pressure $P_\text c$}\\
    \cmidrule(lr){2-6}
    $\xi_\mathrm{s}$ & $\varepsilon^0$ & $\varepsilon^2$ & $\varepsilon^4$ 
           & $\varepsilon^6$ & $\varepsilon^8$ \\ \midrule 
    %%%%%%%%%%%%%%%%
    0.00         	&	0	&	0	&	0	&	0	&	0	\\
0.01      	&	0.00061857	&	-0.00172000	&	0.00162459	&	-0.00077003	&	0.00044127	\\
0.02      	&	0.00243630	&	-0.00686564	&	0.00710374	&	-0.00491391	&	0.00493335	\\
0.03      	&	0.00539833	&	-0.01545281	&	0.01766303	&	-0.01697127	&	0.02537089	\\
0.04      	&	0.00945262	&	-0.02755907	&	0.03514794	&	-0.04585596	&	0.09808811	\\
0.05      	&	0.01454978	&	-0.04334499	&	0.06242121	&	-0.11038522	&	0.34561893	\\
0.06      	&	0.02064301	&	-0.06308734	&	0.10409423	&	-0.25635755	&	1.3551282 	\\
0.07      	&	0.02768796	&	-0.08723187	&	0.16798563	&	-0.63884234	&	4.1589896 	\\
0.08      	&	0.03564262	&	-0.11648064	&	0.26842379	&	-3.0575859 	&	120.34604   	\\
0.09      	&	0.04446728	&	-0.15194212	&	0.43563224	&	0.41855319	&	20.090197  	\\
0.10       	&	0.05412437	&	-0.19540392	&	0.75876395	&	-2.5871083 	&	4.3812886 	\\
0.11      	&	0.06457839	&	-0.24986232	&	2.3812791 	&	-100.52138   	&	7005.1522    	\\
0.12      	&	0.07579584	&	-0.32064223	&	0.88116263	&	-29.802523  	&	-234.29858   	\\
0.13      	&	0.08774509	&	-0.41805092	&	2.3107894 	&	-34.315938  	&	388.81769   	\\
0.14      	&	0.10039634	&	-0.5647382 	&	5.0552036 	&	-87.445248  	&	1669.9741    	\\
0.15      	&	0.11372151	&	-0.82158864	&	12.978500  	&	-348.51779   	&	11069.532     	\\
0.16      	&	0.12769416	&	-1.4259431 	&	52.306181  	&	-3108.6680    	&	224222.47      	\\
0.17      	&	0.14228943	&	-5.0951044 	&	1490.6987    	&	-762280.11      	&	4.84805780
$\times 10^8$	\\
0.18      	&	0.15748397	&	3.0709510 	&	-44.367825  	&	4384.7277    	&	-570869.92      	\\
0.19      	&	0.17325585	&	1.1138325 	&	19.748422  	&	63.519803  	&	-3223.0377    	\\
0.20       	&	0.18958447	&	0.63815819	&	11.158362  	&	84.961966  	&	91.304718  	\\
0.30       	&	0.3798125 	&	-0.11854446	&	1.3921945 	&	5.7408680 	&	18.408063  	\\
0.40       	&	0.61093217	&	-0.28101446	&	0.41114756	&	2.7524237 	&	4.7395486 	\\
0.50       	&	0.8765547 	&	-0.29553405	&	-0.21692694	&	1.7597087 	&	3.1658578 	\\
0.60       	&	1.1744310 	&	-0.19762251	&	-0.64456449	&	0.84355303	&	2.7155097 	\\
0.70       	&	1.5044260 	&	-0.01447731	&	-0.86930239	&	0.02192268	&	1.9510756 	\\
0.80       	&	1.8673640 	&	0.23179265	&	-0.92120251	&	-0.58277995	&	1.0264413 	\\
0.90       	&	2.2644107 	&	0.52592119	&	-0.84400036	&	-0.95208863	&	0.2108996 	\\
1.00         	&	2.6967662 	&	0.85851446	&	-0.67570587	&	-1.1260297 	&	-0.39050992	\\
1.10 	&	3.1655292 	&	1.2242483 	&	-0.44338108	&	-1.1551896 	&	-0.78090453	\\
1.20 	&	3.6716491 	&	1.6203250 	&	-0.16452438	&	-1.0815067 	&	-1.0002778 	\\
1.30 	&	4.2159215 	&	2.0454170 	&	0.15012064	&	-0.93498189	&	-1.0911717 	\\
1.40 	&	4.7990015 	&	2.4990135 	&	0.49416461	&	-0.73571739	&	-1.0879378 	\\
1.50 	&	5.4214241 	&	2.9810379 	&	0.86390299	&	-0.49678082	&	-1.0156282 	\\
1.60 	&	6.0836241 	&	3.4916304 	&	1.2572490 	&	-0.22655363	&	-0.89172864	\\
1.70	&	6.7859549 	&	4.0310276 	&	1.6730773 	&	0.06961668	&	-0.72822437	\\
1.80 	&	7.5287040 	&	4.5994981 	&	2.1108281 	&	0.38832489	&	-0.53328772	\\
1.90 	&	8.3121058 	&	5.1973091 	&	2.5702700 	&	0.72740591	&	-0.31249925	\\
2.00         	&	9.1363524 	&	5.8247100 	&	3.0513580 	&	1.0854878 	&	-0.06968863	\\
    %%%%%%%%%%%%%%%%
   \bottomrule
   \end{tabular}}
   \caption{Expansion coefficients of $\frac{P_{\text c}}{{a_0}^2\,Q^2}$
         for various values of $\xi_\text s$ with $\gamma_i=0$, $i>2$.}
 \end{table}

\begin{table}
  \centerline{
   \begin{tabular}{llllll}\toprule
     & \multicolumn{5}{c}
     {Coefficients of $\varepsilon^i$ for angular momentum $J$}\\
        \cmidrule(lr){2-6}
    $\xi_\mathrm{s}$ & $\varepsilon^0$ & $\varepsilon^2$ & $\varepsilon^4$ 
           & $\varepsilon^6$ & $\varepsilon^8$ \\ \midrule 
    %%%%%%%%%%%%%%%%
    0.00         	&	0	&	0	&	0	&	0	&	0	\\
0.01      	&	0.00519817	&	0.01064142	&	0.01709874	&	0.02545363	&	0.03580801	\\
0.02      	&	0.01452484	&	0.02955774	&	0.04621122	&	0.06832484	&	0.09446591	\\
0.03      	&	0.02637109	&	0.05341626	&	0.08104960	&	0.11945215	&	0.16152228	\\
0.04      	&	0.04014030	&	0.08105972	&	0.11900741	&	0.17572839	&	0.23030652	\\
0.05      	&	0.05548233	&	0.11191984	&	0.15848091	&	0.23645053	&	0.2981537 	\\
0.06      	&	0.07216065	&	0.14575859	&	0.19860914	&	0.30636975	&	0.52287242	\\
0.07      	&	0.09000348	&	0.18260101	&	0.23993267	&	0.43348251	&	-4.1934547 	\\
0.08      	&	0.10888064	&	0.22274743	&	0.28745884	&	1.8835269 	&	-96.001495  	\\
0.09      	&	0.12869081	&	0.26685311	&	0.36460134	&	-2.9069282 	&	-1.4043833 	\\
0.10       	&	0.14935376	&	0.3161068 	&	0.60538199	&	-9.8156498 	&	137.90767   	\\
0.11      	&	0.17080538	&	0.37259919	&	3.8855222 	&	-326.89627   	&	23595.306     	\\
0.12      	&	0.19299422	&	0.44011835	&	-2.1803366 	&	-9.1109666 	&	-1078.5552    	\\
0.13      	&	0.21587897	&	0.52604717	&	-2.3712087 	&	28.000940  	&	-642.86950   	\\
0.14      	&	0.23942672	&	0.64663729	&	-4.3174390 	&	82.390748  	&	-1873.3462    	\\
0.15      	&	0.26361148	&	0.84557362	&	-10.556837  	&	317.16019   	&	-10852.688     	\\
0.16      	&	0.28841317	&	1.2930930 	&	-41.568836  	&	2647.8340    	&	-197589.40      	\\
0.17      	&	0.31381674	&	3.9378673 	&	-1128.4291    	&	585246.47      	&	-3.73910890
$\times 10^8$	\\
0.18      	&	0.33981149	&	-1.9051257 	&	7.2345391 	&	-2888.5244    	&	340738.03      	\\
0.19      	&	0.36639046	&	-0.4890684 	&	-17.726915  	&	-178.76932   	&	-44.181794  	\\
0.20       	&	0.39355005	&	-0.13577661	&	-9.2222335 	&	-105.05367   	&	-807.13712   	\\
0.30       	&	0.69802485	&	0.52311868	&	-0.57132492	&	-4.7290425 	&	-21.880756  	\\
0.40       	&	1.0714436 	&	0.79662582	&	0.05656033	&	-1.1626724 	&	-5.0786338 	\\
0.50       	&	1.5346441 	&	1.0605245 	&	0.24310865	&	-0.220009  	&	-1.5315446 	\\
0.60       	&	2.1149817 	&	1.3709717 	&	0.30021959	&	0.05678398	&	-0.10285   	\\
0.70       	&	2.8447368 	&	1.7637678 	&	0.33048771	&	0.03348169	&	0.45890186	\\
0.80       	&	3.7607272 	&	2.2682180 	&	0.38836636	&	-0.11923423	&	0.53433897	\\
0.90       	&	4.9043081 	&	2.9119034 	&	0.50500367	&	-0.30541288	&	0.35919637	\\
1.00         	&	6.3214817 	&	3.7232237 	&	0.70019629	&	-0.47535233	&	0.07990005	\\
1.10 	&	8.0630209 	&	4.7327036 	&	0.98941174	&	-0.60511924	&	-0.22477905	\\
1.20 	&	10.184575  	&	5.9736140 	&	1.3874168 	&	-0.68287074	&	-0.51722445	\\
1.30 	&	12.746756  	&	7.4822472 	&	1.9099198 	&	-0.70137912	&	-0.78102489	\\
1.40 	&	15.815199  	&	9.2980323 	&	2.5742218 	&	-0.65447365	&	-1.0093842 	\\
1.50 	&	19.460609  	&	11.463577  	&	3.3994232 	&	-0.53552549	&	-1.1992419 	\\
1.60 	&	23.758798  	&	14.024683  	&	4.4064527 	&	-0.33688721	&	-1.3485367 	\\
1.70 	&	28.790709  	&	17.030343  	&	5.6180320 	&	-0.04974338	&	-1.4550308 	\\
1.80 	&	34.642430  	&	20.532741  	&	7.0586270 	&	0.33587825	&	-1.5158600 	\\
1.90 	&	41.405215  	&	24.587246  	&	8.7544002 	&	0.8310431 	&	-1.5274045 	\\
2.00         	&	49.175489  	&	29.252409  	&	10.733173  	&	1.4478377 	&	-1.4852922 	\\
    %%%%%%%%%%%%%%%
    \bottomrule
   \end{tabular}}
   \caption{Expansion coefficients of $\frac{J}{{a_0}^5\,Q^{\nicefrac{3}{2}}}$
         for various values of $\xi_\text s$ with $\gamma_i=0$, $i>2$.}
 \end{table}

\begin{table}
  \centerline{
   \begin{tabular}{llllll}\toprule
     & \multicolumn{5}{c}
      {Coefficients of $\varepsilon^i$ for the radius ratio
	$r_\text p/r_\text e$} \\
    \cmidrule(lr){2-6}
    $\xi_\mathrm{s}$ & $\varepsilon^0$ & $\varepsilon^2$ & $\varepsilon^4$ 
           & $\varepsilon^6$ & $\varepsilon^8$ \\ \midrule 
    %%%%%%%%%%%%%%%%%%%
    0.00         	&	0	&	0	&	0	&	0	&	0	\\
0.01      	&	0.0099995 	&	-0.02602989	&	0.02202631	&	-0.01047418	&	0.00897031	\\
0.02      	&	0.01999600	&	-0.05311731	&	0.05132736	&	-0.03924064	&	0.04708946	\\
0.03      	&	0.02998651	&	-0.08135889	&	0.08911485	&	-0.09392366	&	0.14546497	\\
0.04      	&	0.03996804	&	-0.11089284	&	0.13723154	&	-0.18817272	&	0.37882528	\\
0.05      	&	0.04993762	&	-0.14191637	&	0.19853676	&	-0.3488407 	&	0.96345183	\\
0.06      	&	0.05989229	&	-0.17471238	&	0.27760603	&	-0.63830148	&	2.8459464 	\\
0.07      	&	0.06982913	&	-0.20969134	&	0.38211536	&	-1.2643965 	&	7.3313198 	\\
0.08      	&	0.07974522	&	-0.24746   	&	0.52597493	&	-4.6830624 	&	172.58895   	\\
0.09      	&	0.08963770	&	-0.28893904	&	0.73832306	&	0.09914477	&	26.066958  	\\
0.10       	&	0.09950372	&	-0.33557622	&	1.1055791 	&	-3.4554085 	&	2.2631308 	\\
0.11      	&	0.10934048	&	-0.38975965	&	2.7622215 	&	-101.28858   	&	6954.7987    	\\
0.12      	&	0.11914522	&	-0.45569145	&	1.2805118 	&	-31.530651  	&	-198.73051   	\\
0.13      	&	0.12891523	&	-0.54145696	&	2.6523173 	&	-35.509464  	&	406.54300   	\\
0.14      	&	0.13864784	&	-0.66476338	&	5.1130909 	&	-83.609450  	&	1581.1100    	\\
0.15      	&	0.14834045	&	-0.87312659	&	11.844410  	&	-307.65212   	&	9693.8945    	\\
0.16      	&	0.1579905 	&	-1.3516343 	&	43.829227  	&	-2561.5455    	&	183953.19      	\\
0.17      	&	0.16759549	&	-4.2188500 	&	1173.6956    	&	-598387.12      	&	3.80304840
$\times 10^8$	\\
0.18      	&	0.177153  	&	2.1418255 	&	-41.159648  	&	3581.7333    	&	-454958.98      	\\
0.19      	&	0.18666065	&	0.61071792	&	12.872215  	&	12.113393  	&	-2462.2706    	\\
0.20       	&	0.19611614	&	0.2351988 	&	7.1665406 	&	44.350811  	&	-72.324822  	\\
0.30       	&	0.28734789	&	-0.37867602	&	0.98503149	&	1.8025725 	&	5.8931889 	\\
0.40       	&	0.37139068	&	-0.51715828	&	0.52418984	&	0.79232382	&	0.41160951	\\
0.50       	&	0.4472136 	&	-0.5597459 	&	0.23038891	&	0.69014814	&	0.07257866	\\
0.60       	&	0.51449576	&	-0.55020825	&	0.02038324	&	0.53356119	&	0.27891669	\\
0.70       	&	0.57346234	&	-0.51417099	&	-0.10821692	&	0.34347031	&	0.34737635	\\
0.80       	&	0.62469505	&	-0.46755319	&	-0.17463699	&	0.18619835	&	0.29128276	\\
0.90       	&	0.66896473	&	-0.41919642	&	-0.20141601	&	0.07879986	&	0.20314523	\\
1.00         	&	0.70710678	&	-0.37342997	&	-0.20570353	&	0.01287290	&	0.12721316	\\
1.10 	&	0.73994007	&	-0.33202671	&	-0.19838512	&	-0.0247471 	&	0.07303315	\\
1.20 	&	0.76822128	&	-0.29543502	&	-0.18575195	&	-0.04470894	&	0.03751262	\\
1.30 	&	0.79262399	&	-0.26347544	&	-0.17123161	&	-0.05416609	&	0.01524089	\\
1.40 	&	0.81373347	&	-0.23570866	&	-0.15659797	&	-0.05756331	&	0.00167726	\\
1.50 	&	0.83205029	&	-0.21161961	&	-0.14270895	&	-0.05756365	&	-0.00636027	\\
1.60 	&	0.8479983 	&	-0.19070359	&	-0.12992949	&	-0.05573405	&	-0.01094845	\\
1.70 	&	0.86193422	&	-0.17250244	&	-0.11836672	&	-0.05298648	&	-0.01340201	\\
1.80 	&	0.87415728	&	-0.15661589	&	-0.10799852	&	-0.04984618	&	-0.01454328	\\
1.90 	&	0.88491822	&	-0.14270141	&	-0.09874277	&	-0.04661101	&	-0.01488493	\\
2.00       	&	0.89442719	&	-0.13046906	&	-0.09049408	&	-0.04344527	&	-0.01474681	\\
    %%%%%%%%%%%%%%%%%%%%
   \bottomrule
   \end{tabular}}
   \caption{Expansion coefficients of $r_\text p/r_\text e$
         for various values of $\xi_\text s$ with $\gamma_i=0$, $i>2$.}
 \end{table}
  %%%%%%%%%%%%%%%%%%
 \end{appendix}
 
 \bigskip

\centerline{\bf Acknowledgments}

  Many thanks to R. Meinel, M. Ansorg and J. Ehlers for the valuable
  discussions and helpful advice. 

 \bibliographystyle{myunsrt}
 \bibliography{References}

\end{document}